# Title: Strong tidal dissipation in Saturn and constraints on Enceladus' thermal state from astrometry

**Short title: Tidal dissipation in Saturn from astrometry**

Valéry Lainey[1], Özgür Karatekin[2], Josselin Desmars[3,1], Sébastien Charnoz[4], Jean-Eudes Arlot[1], Nicolai Emelyanov[5,1], Christophe Le Poncin-Lafitte[6], Stéphane Mathis[4], Françoise Remus[7,1,4], Gabriel Tobie[8], Jean-Paul Zahn[7]

[1]*IMCCE-Observatoire de Paris, UMR 8028 du CNRS, UPMC, 77 Av. Denfert-Rochereau, 75014, Paris, France*

[2]*Royal Observatory of Belgium, Avenue Circulaire 3, 1180 Uccle, Bruxelles, Belgium*

[3]*Shanghai Astronomical Observatory, Chinese Academy of Sciences, 80 Nandan Road, 200030 Shanghai, P.R.China*

[4]*Laboratoire AIM, CEA/DSM – CNRS – Université Paris Diderot, IRFU/SAp Centre de Saclay, 91191 Gif-sur-Yvette, France*

[5]*Sternberg Astronomical Institute, 13 Universitetskij Prospect, 119992 Moscow, Russia*

[6]*SyRTE-Observatoire de Paris, UMR 8630 du CNRS, 77 Av. Denfert-Rochereau, 75014 Paris, France*

[7]*LUTH-Observatoire de Paris, UMR 8102 du CNRS, 5 place Jules Janssen, 92195 Meudon Cedex, France*

[8]*Laboratoire de Planétologie et Géodynamique de Nantes, Université de Nantes, CNRS, UMR 6112, 2 rue de la Houssinière, 44322 Nantes Cedex 3, France.*

Corresponding author: lainey@imcce.fr

Abstract: **Tidal interactions between Saturn and its satellites play a crucial role in both the orbital migration of the satellites and the heating of their interiors. Therefore constraining the tidal dissipation of Saturn (here the ratio $k_2/Q$) opens the door to the past evolution of the whole system. If Saturn's tidal ratio can be determined at different frequencies, it may also be possible to constrain the giant planet's interior structure, which is still uncertain. Here, we try to determine Saturn's tidal ratio through its current effect on the orbits of the main moons, using astrometric data spanning more than a century. We find an intense tidal dissipation ($k_2/Q = (2.3 \pm 0.7) \times 10^{-4}$), which is about ten times higher than the usual value estimated from theoretical arguments. As a consequence, eccentricity equilibrium for Enceladus can now account for the huge heat emitted from Enceladus' south pole. Moreover, the measured $k_2/Q$ is found to be poorly sensitive to the tidal frequency, on the short frequency interval considered. This suggests that Saturn's dissipation may not be controlled by turbulent friction in the fluid envelope as commonly believed. If correct, the large tidal expansion of the moon orbits due to this strong Saturnian dissipation would be inconsistent with the moon formations 4.5 Byr ago above the synchronous orbit in the Saturnian subnebulae. But it would be compatible with a new model of satellite formation in which the Saturnian satellites formed possibly over longer time scale at the outer edge of the main rings. In an attempt to take into account for possible significant torques exerted by the rings on Mimas, we fitted a constant rate *da/dt* on Mimas semi-major axis, also. We obtained an unexpected large acceleration related to a negative value of *da/dt*= -(15.7 ± 4.4) × $10^{-15}$ au/day. Such acceleration is about an order of magnitude larger than the tidal deceleration rates observed for the other moons. If not coming from an astrometric artifact associated to the proximity of Saturn's halo, such orbital decay may have significant implications on the Saturn's rings.**

1 Introduction:

Starting with Huygens' observation of Titan in 1655, a little less than two centuries were needed to discover the so-called main moons of Saturn (defined by increasing distance to the primary, Mimas, Enceladus, Tethys, Dione, Rhea, Titan, Hyperion and Iapetus). In common with the Galilean moons, astrometry of the Saturn satellites (consisting of measuring the moon positions in the sky) started in the middle of the $XVII^{th}$ century, with the observations of eclipses by the primary. One has to wait until the end of the $XIX^{th}$ century and the manufacturing of photographic plates, as well as large micrometer and heliometer instruments, for the gathering of reasonably accurate observations (Desmars et al. 2009). Although previously used to probe the gravity fields of the system, astrometry has been replaced advantageously by radio-science data, since the spacecraft era. Nevertheless, the large time span covered by astrometric observations and number of observation sets available can still compensate for any possible lack of precision when one focuses on long term dynamical effects, as for example in the case of Mars (Lainey et al. 2007) and Jupiter (Lainey et al. 2009).

In section 2 we present the observation set used in this study. Section 3 details the numerical model of the Saturnian satellite orbits that has been used to determine Saturn's tidal dissipation. In Sections 4 and 5, we present the fit of the orbit model to astrometric observations and demonstrate its robustness. The last section discusses possible interior models of Saturn and Enceladus in the light of our results.

2 The observation set:

To determine long term effects in the mean motions of the satellites accurately, a set of astrometric observations covering a long time span is necessary. In this context, an extensive catalogue of astrometric observations has been compiled. This catalogue provides about 19 617 observations (counting one date as one observation even for several satellites observed simultaneously) and covers the period from 1886 to 2009. All observations are available in the NSDB natural satellites astrometric database (Arlot, Emelyanov 2009).

The main source of the catalogue is COSS08 (Desmars et al. 2009) yielding about 130 000 data (counting one coordinate of one satellite as one datum) from 1874 to 2007. For our set of data, only the « accurate » observations have been selected. To define « accurate » observations, we have first excluded those with a residual larger than 2 arcsec, then we have computed the root mean square (rms) of the residuals for all the observations corresponding to each bibliographic reference. Finally, for the purposes of the present paper, if the rms was larger than 0.3 arcsec for observations before 1950 and 0.25 arcsec for observations after 1950, the entire set issued from that particular bibliographic reference has been excluded. As a consequence, about 93 % of COSS08 observations have been selected starting from 1886 until 2007.

Since 2007, other observations have become available. The USNO Flagstaff transit circle data, already included in COSS08, has been updated until 2009. Observations from (Peng et al. 2008) have also been added. Moreover the highly accurate astrometric observations provided by the observation of the mutual phenomena of Saturnian satellites during 1995 and 2009 have been added.

Finally, the extensive catalogue contains about 19 000 observations made from 1886 to 2009.

Direct astrometric observations can be performed in several different ways: transit observations, photographic plates and CCD imaging. They are reduced from the known position of reference stars visible in the field of view during the observation. Because of their small field of view, the CCD frames and some long focus photographic plates do not often allow the use of reference stars. To deal with such data, the authors generally use the position of specific well known satellites (usually Titan, Rhea, Dione and Tethys because of their accurate ephemerides) computed with a specific theory as a reference in order to deduce the astrometric positions of the other satellites. A drawback of this is that these observations may be biased by any limiting assumptions in the adopted theory. To deal with this problem, we preferred to consider only relative separation and position angle between the satellites using pixel positions. This method provides astrometric observations that do not depend on the theory. It has been applied for CCD observations when stars were not used in the astrometric reduction. The influence of such a bias is tested in subsection 5.3.

Photometric observations of mutual occultations and eclipses of the Saturnian satellites provide very accurate astrometric relative positions of the satellites. These observations are possible during the Saturnian equinox, since the Sun and the Earth are then in the equatorial plane of Saturn which is also the common orbital plane of the satellites. Campaigns of observations of such mutual occultations and eclipses were made in 1995 and 2009. We undertook the processing of the complete database of these photometric observations published by (Thuillot et al. 2001). An accurate photometric model of mutual events using the scattering properties of the satellite surfaces (Buratti 1984) issued from Voyager data was used. A large analysis of the properties of saturnian icy satellites is done in (Pitman et al. 2010) using the observations of these bodies provided by the Cassini probe. Phase curves are given in the V and R bands for solar phase angles in the vaste range up to 180 degrees (but only for Rhea and Dione). Unfortunately, because of this wide range of phase angles, we could not rely on (Pitman et al. 2010) for the satellite albedo in the range of 1 to 6 degrees and therefore did not use their data.

In order to extract astrometric positions from photometric data, we developed an original method (Emelyanov & Gilbert 2003). We have processed the 46 light curves obtained during the international campaign of photometric observations of the Saturnian satellites in 1995 and the 17 light curves obtained during the international campaign in 2009. From these photometric observations 46 topocentric or heliocentric angular differences in right ascension and declination for satellites pairs on the time interval from December 16, 1995 to February 6, 1996 and 17 topocentric or heliocentric angular differences on the time interval from December 19, 2008 to July 7, 2009 were obtained. The errors due to random errors of photometry are from 1 to 15 mas in right ascension and declination and characterize the internal accuracy of the astrometric results. Nevertheless, due to the rather small number of observed events compared to other observation sets, the contribution of mutual events has turned out to be modest. This contrasts with the Jovian case (Lainey et al. 2009).

3 The dynamical model:

The NOE (Numerical Orbit and Ephemerides) numerical code (Lainey et al. 2007; Lainey 2008; Lainey et al. 2009) has been used to model the orbital motion of the Saturnian satellites. It is a gravitational N-body code that incorporates highly sensitive modeling and can generate partial derivatives needed to fit initial positions, velocities, and other parameters (like the ratio $k_2/Q$) to the observational data. The code includes (i) gravitational interaction up to degree two in the spherical harmonic expansion of the gravitational potential for the satellites and up to degree 6 for Saturn with the numerical values from (Jacobson et al. 2006); (ii) perturbations due to the Sun and Jupiter using DE406 ephemerides (with the inner planets and the Moon included by incorporating their masses in the Solar value); (iii) Saturnian precession from (Jacobson 2007); (iv) tidal effects introduced by means of the Love number $k_2$ and the quality factor Q in the combination $k_2/Q$ for Saturn and Enceladus. The orbital effects due to the dissipation inside Saturnian satellites other than Enceladus are neglected, since they are expected to be much less dissipative, less eccentric or much further away from Saturn. Nevertheless, the tidal bulges raised by each moon on Saturn are taken into account.

The dynamical equations are numerically integrated in a Saturncentric frame with inertial axes (conveniently the Earth mean equator J2000). The equation of motion for a satellite $P_i$ can be expressed in a general form as

$$\ddot{\vec{r}}_i = -G(m_0 + m_i)\left(\frac{\vec{r}_i}{r_i^3} - \nabla_i U_{\bar{i}\hat{0}} + \nabla_0 U_{\bar{0}\hat{i}}\right) + \sum_{j=1, j\neq i}^{N} Gm_j \left(\frac{\vec{r}_j - \vec{r}_i}{r_{ij}^3} - \frac{\vec{r}_j}{r_j^3} - \nabla_j U_{\bar{j}\hat{i}} + \nabla_i U_{\bar{i}\hat{j}} + \nabla_j U_{\bar{j}\hat{0}} - \nabla_0 U_{\bar{0}\hat{j}}\right)$$

$$+ \frac{(m_0 + m_i)}{m_i m_0}\left(\vec{F}_{\bar{i}\hat{0}}^T - \vec{F}_{\bar{0}\hat{i}}^T\right) - \frac{1}{m_0}\sum_{j=1, j\neq i}^{N}\left(\vec{F}_{\bar{j}\hat{0}}^T - \vec{F}_{\bar{0}\hat{j}}^T\right) \quad (1)$$

Here, $\vec{r}_i$ and $\vec{r}_j$ are the position vectors of the satellite $P_i$ and a body $P_j$ (another satellite, the Sun, or Jupiter) with mass $m_j$, subscript 0 denotes Saturn, $U_{\vec{kl}}$ is the oblateness gravity field of body $P_l$ at the position of body $P_k$ and $\vec{F}_{\vec{lk}}^{\,T}$ the force received by $P_l$ from the tides it raises on $P_k$. This force is equal to (Lainey et al. 2007):

$$\vec{F}_{\vec{lk}}^{\,T} = -\frac{3k_2 G m_l^2 R^5 \Delta t}{r_{kl}^8}\left(\frac{2\vec{r}_{kl}(\vec{r}_{kl}\cdot\vec{v}_{kl})}{r_{kl}^2} + (\vec{r}_{kl}\times\vec{\Omega}+\vec{v}_{kl})\right) \quad (2)$$

where $\vec{r}_{kl} = \vec{r}_k - \vec{r}_l$, $\vec{v}_{kl} = d\vec{r}_{kl}/dt$, $\vec{\Omega}$, $R$, and $\Delta t$ being the instantaneous rotation vector, equatorial radius and time potential lag of $P_k$, respectively. The usual tidal term independent of Q (and so only dependent on $k_2$) that arises in the tidal potential development has been neglected here. This is justified for two reasons: it is pretty small (a typical drift of a few tens of km in longitude after 100 years); and most importantly, since it provides only secular drift (but not secular acceleration) on longitudes, it can be easily absorbed in a tiny change of the initial conditions without any significant consequences. While it was considered only for completeness in (Lainey et al. 2009), it has been neglected here.

The time lag $\Delta t$ is defined by (Lainey et al. 2007)

$\Delta t = T\,\mathrm{Atan}(1/Q)/2\pi$ \quad (3)

where $T$ is the period of the main tidal excitation. For the tides raised on Enceladus, $T$ is equal to $2\pi/n$ ($n$ being Enceladus' mean motion) as we only considered the tide raised by Saturn. For Saturn's tidal dissipation, $T$ is equal to $2\pi/2(\Omega-n_i)$ where $\Omega$ is the spin frequency of Saturn and $n_i$ is the mean motion of the tide raising Saturnian moon $P_i$. $\Delta t$ depends on the tidal frequency and on Q, therefore it is not a constant parameter.

It is clear from the second term in the right hand side of equations (2-3) that $k_2$ and Q are completely correlated. In practice, we considered the commonly used value $k_2$=0.341 (Gavrilov & Zharkov 1977) and fitted only Q.

Because of a 2:1 resonance located at the outer edge of the B-ring, Saturn's rings are expected to interact significantly with Mimas (Lissaueur & Cuzzi 1982). However the magnitude of this effect is unknown, because of large uncertainties about the ring structures and surface densities. To take into account such an interaction, we had to introduce a supplementary force in the system modelling a constant rate *da/dt* on Mimas' semi-major axis (denoted *a*), and considered as an additional free parameter in the model. As a consequence, no information on tidal dissipation inside Saturn may be obtained directly from Mimas' orbital motion, since this latter is mixed with the estimation of the ring dynamical effects. Moreover, because of the Mimas-Tethys resonant interaction, such a *da/dt* rate should not be compared with a possible observed kinematic rate. To introduce a *da/dt* constant term as a supplementary force in the model, we used the Gauss equations. We recall that this differential system provides the variation of Keplerian elements as a function of disturbing forces expressed in the local base. Introducing a constant variation in the semi-major axis (and no variations in the other Keplerian elements), this system can be easily inverted to provide the proper expression of the force.

For an unspecified parameter $c_l$ of the model that shall be fitted (e.g. $\vec{r}(t_0)$, $d\vec{r}/dt(t_0)$, Q…), a useful relation is

$$\frac{\partial}{\partial c_l}\left(\frac{d^2\vec{r}_i}{dt^2}\right) = \frac{1}{m_i}\left[\sum_j\left(\frac{\partial \vec{F}_i}{\partial \vec{r}_j}\frac{\partial \vec{r}_j}{\partial c_l} + \frac{\partial \vec{F}_i}{\partial \dot{\vec{r}}_j}\frac{\partial \dot{\vec{r}}_j}{\partial c_l}\right) + \frac{\partial \vec{F}_i}{\partial c_l}\right], \qquad (4)$$

where $\vec{F}_i$ is the right hand side of Eq.(1) multiplied by $m_i$.

Partial derivatives of the solutions with respect to initial positions and velocities of the satellites and dynamical parameters are computed from simultaneous integration of equation (4) and equation (1). For an explicit formulation of the dynamical equations and the variational equations used, we refer to (Lainey et al. 2007; Lainey 2008; Lainey et al 2009) (and references therein).

The RA15 numerical integrator is used with a constant step size of 0.075 day. To increase the numerical accuracy, we performed forward and backward integrations starting at an initial Julian epoch of 2433291.0 (9[th] of January 1950 TDB). The numerical accuracy of our simulation is at the level of a few hundreds of metres over the whole 123 years (see also Appendix A.1.1).

During the fitting procedure, time scale and light time corrections for each satellite-observer distance were introduced (Lainey et al. 2007). Corrections for phase, aberration and differential refraction were applied when they were not already included in the observation astrometric reductions (Lainey et al. 2007). Observational subsets (related to different observational campaigns or publications) have been considered with a relative weight computed by preliminary residuals (Lainey et al. 2007) and corresponding to their root mean square error for deriving formal errors. Least-squares iterations have been applied to solve for the fitted parameters. In particular, despite the development of new techniques, least squares is still one of the most efficient method available to solve for the parameters of dynamical systems (Desmars et al. 2009b), as long as the studied system has been observed over a sufficiently long period of time to allow for rather accurate initial conditions (which is the case for the main planetary satellites of the Solar system). No constraints have been applied in the least squares inversion. Only a few iterations have been required to reach an optimal solution.

In all this work, the fitted parameters are the initial state vectors of the main Saturnian moons (actually their equivalent form as Keplerian elements) of all Saturnian moons, the extra parameter da/dt for Mimas and the ratio $k_2/Q$ for Saturn. In particular, Enceladus' tidal dissipation could not be fitted due to significant correlations. To solve this issue, we considered two extreme scenarios for each solution: i) Enceladus is at thermal equilibrium; ii) Enceladus is not dissipative at all. Then we merged both solutions into one, providing one global solution with higher error bars, but independent of Enceladus' internal state (see Section 4). Hence, the total number of fitted parameters considered is between 50 and 53 (depending on whether Q is assumed constant or dependent on the tidal frequency).

4 Fitting the model to astrometric observations:

The dependence of Q on tidal frequency is a matter for debate. While it is traditionally approximated by a constant for long time scales (Goldreich & Soter 1966; Sinclair 1983), recent developments in the numerical simulation of giant planet interiors has revealed a possible erratic frequency-dependence of Q (Wu 2005). In this work we investigate both possibilities.

### 4.1 Constant Q model:

In a first inversion, we neglect dissipation in Enceladus and fit the initial state vectors of all eight moons including Saturn's ratio $k_2/Q$ and Mimas' $da/dt$. We thus obtain $k_2/Q = (2.0 \pm 0.4) \times 10^{-4}$ and $da/dt=-(13.7 \pm 2.4) \times 10^{-15}$ au/day. Saturn's dissipation ratio corresponds to orbital acceleration values $\dot{n}/n$ (in yr$^{-1}$ units) of -(2.67 ± 0.57) × 10$^{-10}$, -(4.26 ± 0.91) × 10$^{-10}$, -(1.52 ± 0.33) × 10$^{-10}$ and -(3.56 ± 0.76) × 10$^{-11}$ for Enceladus, Tethys, Dione and Rhea, respectively. This translates in semi-major axis variation $da/dt$ (in au/day units) of (7.77 ± 1.67) × 10$^{-16}$, (1.53 ± 0.33) × 10$^{-15}$, (7.02 ± 1.50) × 10$^{-16}$, (2.29 ± 0.49) × 10$^{-16}$. Over the 123 years covered by the observation set we used, this corresponds to an orbital shift in longitude of 799 ± 172 km (0.129 ± 0.028 arcsec), 1152 ± 246 km (0.186 ± 0.040 arcsec), 365 ± 78 km (0.059 ± 0.013 arcsec), 72 ± 15 km (0.012 ± 0.002 arcsec), respectively. In a second case, we introduce dissipation in Enceladus, which is expected to counterbalance its orbital acceleration, thereby to modifying our global estimation of Saturn $k_2/Q$ obtained from the satellite tidal accelerations. As already stated in Section 3, we do not have a sufficient number of observations to invert independently the $k_2/Q$ values for Saturn and Enceladus. Hence, to introduce Enceladus' tidal dissipation we assume that Enceladus is in a dynamical equilibrium state, which locks Enceladus' eccentricity as the result of dissipation in both Saturn and Enceladus inside the 2:1 resonance with Dione (Meyer & Wisdom 2007). In this case, we obtain $k_2/Q = (2.6 \pm 0.4) \times 10^{-4}$ and $da/dt=-(17.0 \pm 2.4) \times 10^{-15}$ au/day. The associated secular accelerations related to Saturn's and Enceladus' tides are - (2.06 ± 0.57) × 10$^{-10}$, -(5.61 ± 0.91) × 10$^{-10}$, -(2.09 ± 0.32) × 10$^{-10}$, -(4.69 ± 0.76) × 10$^{-11}$. This translates in semi-major axis variation $da/dt$ (in au/day units) of (6.00 ± 1.66) × 10$^{-16}$, (2.02 ± 0.33) × 10$^{-15}$, (9.65 ± 1.49) × 10$^{-16}$, (3.02 ± 0.49) × 10$^{-16}$. This corresponds to 619 ± 171 km (0.100 ± 0.028 arcsec), 1516 ± 245 km (0.245 ± 0.040 arcsec), 501 ± 78 km (0.081 ± 0.013 arcsec) and 95 ± 15 km (0.015 ± 0.002 arcsec), also. Even though both inversions provide the same orbital trends, acceleration values for each satellite are somewhat different. This arises from the use of only one global $k_2/Q$ value for Saturn, while fitting several independent accelerations. Thanks to the long time span considered, astrometric accuracy is enough to detect tidal accelerations for Enceladus, Tethys, Dione and Rhea from the observations (see Figure 1 and Tables 1, 2 and 3). Combining the two fits, our nominal solution for the Saturn tidal dissipation (here assumed to be independent on the tidal frequency) yields $k_2/Q= (2.3 \pm 0.7) \times 10^{-4}$.

4.2 Non-constant Q model:

To check the assumption of a constant Q model considered in our nominal solution, we release simultaneously several Q values (with and without Enceladus' dissipation), one Q value being related to each tide raising satellite. We succeed in obtaining Saturn's Q at four different tidal frequencies, related to Enceladus, Tethys, Dione and Rhea (see Figure 2), respectively. Correlations between the four tidal ratios are below 0.2 (as shown in Table 4). However, a high correlation of 0.935 is found between Saturn's tidal ratio associated with Tethys' tidal frequency and Mimas' *da/dt* (equal to $-(16.3 \pm 3.7) \times 10^{-15}$ au/day), as a consequence of the 2:4 mean motion resonance between Mimas and Tethys.

Implications of these results are discussed in Section 6.

5 Robustness of the solution:

In this section we present various tests that assess the robustness of our solution. During these tests, we considered as a reference solution the Enceladus equilibrium scenario with a constant Saturn Q model, that is $k_2/Q= (2.6 \pm 0.4) \times 10^{-4}$ and $da/dt=-(16.9 \pm 2.4) \times 10^{-15}$ au/day.

5.1 Random holdout method:

This method considers the change in fitted parameters when a constant percentage of observations is removed. We have performed a test, with a percentage of 10% of observations that are not used. The number of observations in the full nominal least squares inversion is 19616 (all moon coordinates at a given time considered as one observation). 100 different subsets were generated and used to check the robustness of the fitted parameters. We obtained the $k_2/Q$-value in the interval [$1.8 \times 10^{-4}$, $3.3 \times 10^{-4}$] for Saturn and the Mimas da/dt value in the interval [$-21.9 \times 10^{-15}$, $-9.1 \times 10^{-15}$] au/day (even though most values were in agreement with the nominal error bars). This suggests that factors of 2 and 4 could be introduced in our nominal error bars in $k_2/Q$ and da/dt, respectively. However, it must be remembered that Mimas and Tethys are hard to observe from the ground. Hence, significantly decreasing the number of observations of Mimas will automatically increase the chance of losing the da/dt signal.

5.2 Removing successively the five largest observation subsets:

In this test, we completely removed several observation subsets among the most dense ones: i) Vienne et al. 2001; ii) Vass (1997); iii) FASTT observations (see Stone & Harris 2000 and references therin); iv) Struve (1898); v) USNO (1929). The first three observation subsets are associated with the modern era, while the last two consist of observations from the end of the XIX$^{th}$ century and beginning of the XX$^{th}$ century respectively. We provide in Table 5 the fitted value of $k_2/Q$ and da/dt after having removed the mentioned subset.

With the exception of removing the Struve data, all solutions above are very close to (though not in full agreement with) our nominal solution. The Struve data even indicate a slightly higher tidal dissipation ratio. We conclude that removing any observational subsets (old or modern) still confirms the high Saturnian dissipation obtained in our nominal solution.

5.3 On the use of pixel positions:

As already mentioned in section 2, the ephemerides of the outermost Saturnian moons are sometimes used to determine the scale and orientation of the observations (because of the lack of stars in the observed fields). This can be justified since outermost moons (Dione, Rhea, Titan and sometimes Iapetus) are easier to observe (entailing a more accurate ephemeris). Moreover, an error on the scale factor and orientation is expected to have a smaller influence on the innermost moons that are close to the center of the observation. Nevertheless, introducing possibly an external orbital model in the reduction of the observations is not fully satisfactory. This is why we decided, when available, to use pixel positions instead of $(\alpha,\delta)$ or $(s,p)$ coordinates. Under this form, the observations are not corrupted by any external dynamical model, but their significance in the fit is lower, since information on the scale and orientation is no longer present.

We have checked the difference between our nominal solution (using pixel positions) and a fit using the usual astrometric coordinates $(\alpha,\delta)$ or $(s,p)$. We obtained the following result: $k_2/Q = (1.8 \pm 0.4) \times 10^{-4}$ for Saturn and $da/dt = -(13.0 \pm 2.4) \times 10^{-15}$ for Mimas. This is in agreement with our nominal solution (taking into account the error bars).

5.4 Scale factor biases:

One of the crucial systematic errors in astrometric observations is related to scale factors. Scale factors express the equivalence between an observed distance on a field

(measured in micrometer on a photographic plate, or pixel on a CCD image) and its related angular separation on the celestial sphere. In principle, a scale factor should be an isotropic quantity. Nevertheless, stellar positions used to calibrate the observations are rarely corrected for atmospheric differential refraction. Hence, scale factors along equatorial and polar directions are different. Most of the time, old observations used a constant scale factor, which introduced systematic errors in the satellite positions. In particular, such errors produce higher residuals for the distant satellites than for the closer ones, which are much less affected by a small error in the field scale.

To check the influence of scale factor errors on our results, we have performed another fit of our model using this time the relative distance of the satellites between each other (instead of the absolute distance derived from the scale factor estimation), when at least 3 satellites were observed simultaneously. We obtained the following result: $k_2/Q= (2.7 \pm 0.3) \times 10^{-4}$ for Saturn and $da/dt=-(17.9 \pm 1.8) \times 10^{-15}$ for Mimas. These two values agree with our nominal solution within the error bars.

5.5 Releasing more parameters in the fit (Saturn's pole and precession):

While the gravity field parameters we are using are accurate thanks to Cassini data, the IAU expression for Saturn's pole coordinates and precession frequency dates back to 1994 (Davies et al. 1996; Archinal et al. 2011). Here we investigate the influence of fitting the pole coordinates and precession frequency of Saturn on our results. Starting from our nominal solution (the Enceladus equilibrium model), we performed a new fit that added four more parameters in the fitting process. These parameters refer to the IAU formulation for the pole coordinates of planets (Archinal et al. 2011):

$$\alpha = \alpha_0 + \dot{\alpha}T$$
$$\delta = \delta_0 + \dot{\delta}T$$

where ($\alpha_0, \delta_0$) are the pole coordinates in the ICRF at epoch J2000.0 and ($\dot{\alpha}, \dot{\delta}$) introduces the precession/nutation of the primary. Releasing the 50+4 parameters simultaneously, we obtain after iterations the following solution: $k_2/Q$= (2.63 ± 0.41) × $10^{-4}$, da/dt= -(16.7 ± 2.4) × $10^{-15}$ au/day, $\alpha_0$= 40.5915 ± 0.0055 deg, $\dot{\alpha}$ = -0.131 ± 0.022 deg/Julian century, $\delta_0$= 83.54163 ± 000053 deg, $\dot{\delta}$ =0.0219 ± 0.0025 deg/Julian century.

Clearly $k_2/Q$ and da/dt are poorly affected by adding to the fit Saturn's pole coordinates and precession frequency. Moreover, the post fit residuals are highly similar to those from the nominal solutions (in the limit of one tenth of a mas). Hence, our results are expected to be poorly sensitive to possible IAU errors in the expression of Saturn's pole coordinates and precession frequency.

5.6 Neglecting general relativity:

Here we focus on the influence of general relativity. Since the influence of these effects is pretty small (Iorio & Lainey 2005*)*, they have been neglected in our nominal fits. To demonstrate the validity of such an assumption, we have introduced these effects in the model (we considered relativistic effects associated to both Saturn and the Sun) and performed a new solution. After fitting, we obtained: $k_2/Q$= (2.54 ± 0.42) × $10^{-4}$, da/dt= -(16.9 ± 2.4) × $10^{-15}$ au/day.

Here again, the post fit residuals are highly similar to the nominal solutions ones (in the limit of one tenth of a mas).

In conclusion, the envelope of all the tests performed in this whole section provides $k_2/Q = (3.1 \pm 1.7) \times 10^{-4}$ and $da/dt = -(16.2 \pm 7.6) \times 10^{-15}$ au/day. In particular, our different tests confirm that the present estimation of the $k_2/Q$ ratio of Saturn is reliable.

6 Discussion:

We can see from subsection 4.2 that all $k_2/Q$ values lie in the same range and show smooth frequency dependence. To understand the implications of this result, let us recall what is presently known about the physical mechanisms that convert into heat the kinetic energy of the tides, thus driving the secular evolution of the system. Saturn has a hydrogen-helium fluid envelope, which may be partially or entirely convective, and an expected rock-ice core (Guillot 1999) (ignoring here the role of a possible outer radiative layer). The fluid equilibrium tide is damped mainly in the convective part of the envelope and its amplitude depends smoothly on frequency (Zahn 1966; Zahn 1989). In contrast, the dynamical tide in this region, which consists of excited inertial modes (Ogilvie & Lin 2004; Wu 2005; Goodman & Lackner 2009), varies considerably with frequency, particularly in the presence of a dense core (Rieutord & Valdettaro 2010). Although both of these fluid tides are damped through turbulent friction, they yield a relatively low value of the dissipation parameter: $k_2/Q \sim 10^{-6}$ at most (Ogilvie & Lin 2004). Therefore one has to turn to other mechanisms to explain the high tidal dissipation in Saturn's system that we report here, with $k_2/Q$ ranging from $1.4 \times 10^{-4}$ to $3.4 \times 10^{-4}$. For instance, the contribution to tidal dissipation of a stably stratified layer surrounding the core, due to the settling of helium when it ceases to be soluble in metallic hydrogen (Morales et al. 2009; Fortney & William 2003), remains to be evaluated. Another possibility could be the presence of a dense core, as predicted by most models (Guillot 1999). In Saturn this core is expected to be relatively larger than in Jupiter, which would be consistent with the relatively lower tidal dissipation in that planet: $k_2/Q = (1.102 \pm 0.203) \times 10^{-5}$, as determined by the same method (Lainey et al. 2009).

So far, the averaged lower bound of Saturn's Q was derived from theoretical considerations, assuming that all main moons formed above the synchronous orbit 4.5 Byr ago (Goldreich & Soter 1966). Considering Mimas, the innermost mid-sized satellite, and using the averaged equations for a tidally evolving system, (Sinclair 1983) derived a present-day reference value of Q≥18,000 (assuming $k_2$=0.341 from (Gavrilov & Zharkov 1977)). If the observed high value of Saturn's $k_2/Q$ determined in this work represents well the long-term averaged value, then a very large tidal expansion of the moon orbits should have occurred. In particular, the conventional assumption of Saturnian satellites forming contemporaneously with their parent planet has to be dismissed. Recently, Charnoz et al. (2011) suggested a new mechanism of formation of Saturnian satellites at the outer edge of the rings. While the satellites evolve outward in their model due to exchange of angular momentum with the rings and tides rose in the primary, they noticed however that a strong tidal dissipation in Saturn is mandatory to place the satellites at their current observed positions. It is noteworthy that the tidal dissipation quantification presented here allows their model to form and place the Saturnian satellites at the proper distance to their primary.

Since the discovery of the very active province at Enceladus' south pole by the Cassini spacecraft in 2005 (Spencer et al. 2009; Porco et al. 2006), a variety of theoretical models have been proposed to explain the huge thermal emission as well as the associated eruptions of water vapour and ice particles (Meyer & Wisdom 2007; Nimmo et al. 2007; Tobie et al. 2008; O'Neill & Nimmo 2010; Howett et al. 2011). The amount of energy produced at present by radioactive decay in the rocky core of Enceladus contributes less than 2% to the total emitted power (15.8 ± 3.1 GW (Howett et al. 2011), suggesting another internal energy source such as tidal dissipation. Previous studies based on the former estimation of Saturn's Q

(≥18,000) (Sinclair 1983) showed that tidal heating at orbital equilibrium, required to maintain the resonant Enceladus-Dione orbital configuration, could not account for more than 1.1 GW (Meyer & Wisdom 2007). This was suggesting that either the resonant system oscillates around equilibrium with dissipated power varying from almost zero to the observed value or more, or the satellite episodically releases the internal heat that is continuously produced at a rate compatible with orbital equilibrium (Tobie et al. 2008; O'Neill & Nimmo 2010). None of these solutions was fully satisfactory.

The new dissipation factor of Saturn reported here totally changes our understanding of Enceladus' heat production mechanism. As shown on Figure 3, with the new inferred Q value, equilibrium tidal heating can now account for the observed heat power. This indicates that Enceladus' interior could be close to thermal equilibrium at present, surface heat loss being balanced by heat produced by tidal dissipation. Large tidal dissipation in Enceladus implies that the satellite probably possesses a liquid water layer decoupling the outer ice shell from the rocky core (Nimmo et al. 2007; Tobie et al. 2008). For Saturnian Q lower than 2000, Enceladus can remain highly dissipative during a very long period of time without damping its orbital eccentricity, and the long-term stability of a subsurface ocean would thus be possible.

7 Conclusion:

We have quantified Saturn's tidal dissipation ratio $k_2/Q$ to be $(2.3 \pm 0.7) \times 10^{-4}$ using astrometric observations spanning 123 years. Moreover, such a quantification directly derived from observations is provided here at various frequencies, for the first time in a giant planet. As a consequence, we conclude that tidal dissipation may mostly occur in Saturn's core and its boundary. Moreover, tidal heating equilibrium is now a possible state for Enceladus.

The present quantification of Saturnian tidal dissipation is incompatible with a satellite formation scenario in the Saturn's subnebulae for all moons below Titan. However, it is fully compatible with a formation at the edge of Saturn's rings (Charnoz et al. 2011).

During all fitting procedures, we obtain an extra acceleration on Mimas' orbital longitude, related to a negative value of $da/dt$= -(15.7 ± 4.4) × $10^{-15}$ au/day (combination of all fit values). A possible source of error explaining such decay could be the proximity of Mimas to Saturn's halo. But if confirmed in the future, Mimas orbital decay could have significant implications on the Saturn's rings.

Acknowledgments:

This work has been supported by EMERGENCE-UPMC grant (contract number: EME0911), FP7 program Europlanet-RI, EC grant (228319) and PNP (CNRS/INSU). The authors are indebted to all participants of the French Encelade WG and to Chloé Michaut. V.L. would like to thank Nick Cooper, Carl Murray, Michael Efroimsky, Alain Vienne and Laurène Beauvalet for fruitful discussions.

Appendices:

A1 Testing our numerical model:

It is not possible to show the very large amount of tests that have been performed over years when developing the NOE code. Nevertheless, in the following section we provide several tests that can be considered as fundamental to our study.

Unless explicitly mentioned, the tested model here introduces the Enceladus equilibrium scenario with constant Q.

A.1.1 Numerical precision:

To test the numerical precision of our integrations of the equations of motion, we performed backward and forward integrations (see also the Conservation of Energy test). The Figure A1 below shows the difference, for each moon, between backward and forward integration expressed as Euclidian distance in km, over one century. These variations are all below 400 meters. Since our integrations have been performed over no longer than 64 years (we recall that our fit epoch is 1950, and we cover the period 1886-2009), we can conclude that the numerical precision of the satellite positions in our study is a few hundreds of meters.

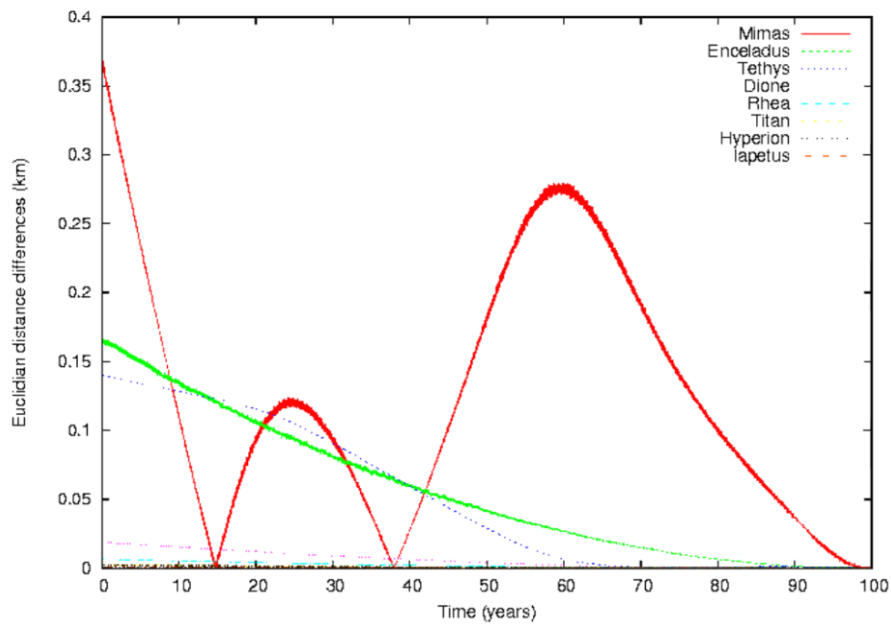

A.1.2 Conservation of energy:

Checking the conservation of energy is quite important for two reasons:

1- It provides another way to quantify numerical error (here the numerical accuracy)
2- More importantly, it checks the validity of the force model

Nevertheless, energy is not always conserved for "any" force model (non conservative forces, use of planetary ephemerides, introduction of forced precession and spin-orbit coupling…). In the following test (see Figure below), we did not introduce i) da/dt "force", ii) tidal dissipation, iii) Saturn's precession. Moreover, Solar and Saturnian motions have been integrated explicitly (no use of planetary ephemerides) and the rotation of the moons has been frozen.

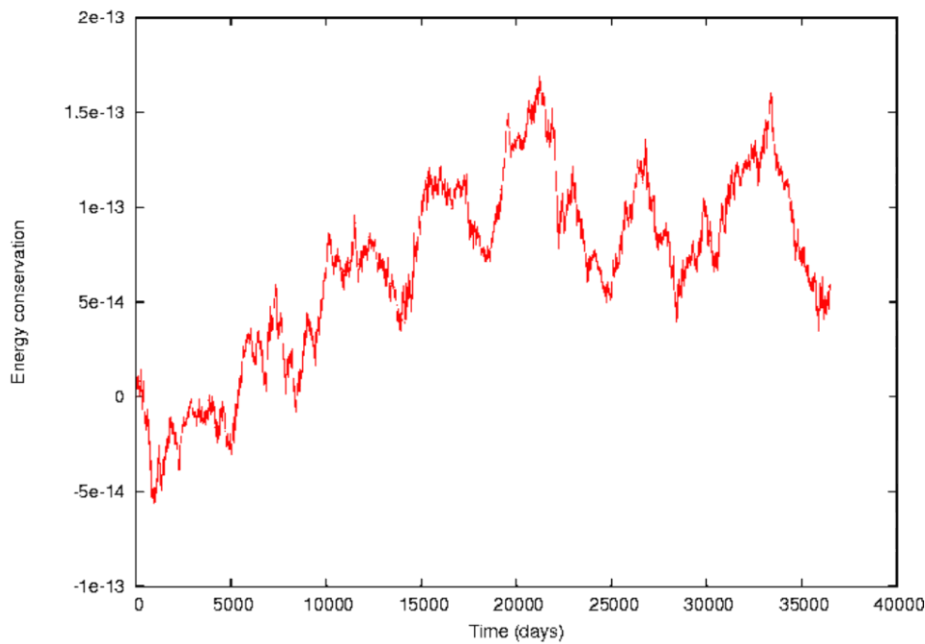

As one can see on Figure A2, energy is conserved up to 13 digits (use of double precision) over 100 years, in our model.

A.1.3 Testing the tidal model:

Since conservation of energy cannot be used to check the validity of our tidal model, we have used analytical expression of da/dt and de/dt to check the code. In particular, we recall that we have (as a first approximation) for the tides raised in the primary (Kaula 1964):

$$\frac{da}{dt} = \frac{3k_2 mnR^5}{QMa^4}\left(1 + \frac{51}{4}e^2\right)$$
$$\frac{de}{dt} = \frac{57k_2 mn}{8QM}\left(\frac{R}{a}\right)^5 e \qquad (A1)$$

and for the tides raised in the 1:1 spin-orbit satellite (Peale & Cassen 1978):

$$\frac{da}{dt} = -\frac{21k_2^s MnR_s^5}{Q^s ma^4}e^2$$
$$\frac{de}{dt} = -\frac{21k_2^s Mn}{2Q^s m}\left(\frac{R_s}{a}\right)^5 e \qquad (A2)$$

To make the comparison straightforward, we first considered a two-body problem for each moon. However, integrating the Saturnian system by modelling only 2-body interactions requires considering a different eccentricity when using the equations (A1-A2). In particular, we have found that using our nominal solution as initial conditions, the eccentricity of each moon was changed to 0.017011, 0.00534, 0.000976, 0.00188, 0.00114, and 0.02895 for Mimas to Titan respectively.

A.1.3.1 Tides in the planet (2-body problem):

Over 100 years, variations on $a$ are expected to be 8.37, 6.08, 10.722, 4.889, 1.641, 0.952 meters for respectively S1, ..., S8 (Mimas... Titan). Similarly, variations on $e$ are expected to be $1.81 \times 10^{-9}$, $3.24 \times 10^{-10}$, $8.46 \times 10^{-11}$, $5.78 \times 10^{-11}$, $8.42 \times 10^{-12}$, $5.30 \times 10^{-11}$. Comparing such estimations with our numerical simulation offers a good agreement (see Figures A3-A4; numerical table is available on request).

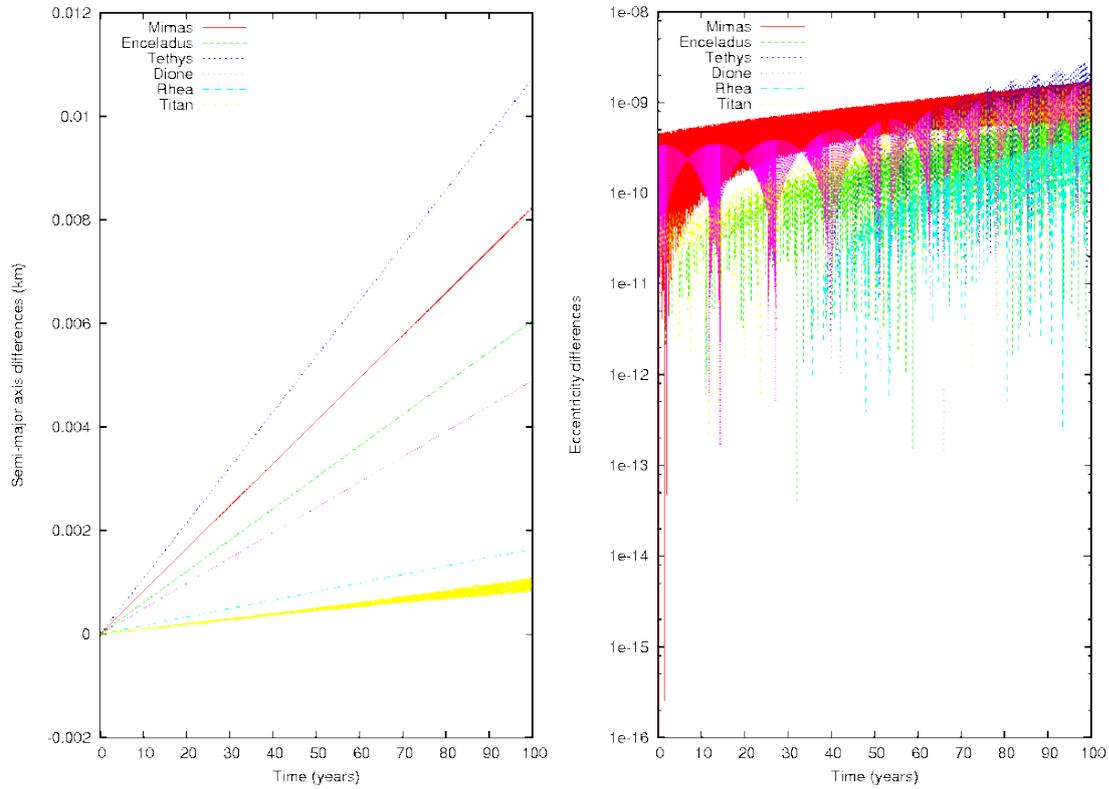

A.1.3.2 Tides in the satellites (2-body problem):

Over 100 years and assuming arbitrarily $k_2^s/Q^s=10^{-2}$ for all moons, variations on *a* are expected to be -54.607, -1.597, -0.124, -0.087, -0.012, -0.544 meters for respectively S1, ..., S8 (Mimas... Titan). Similarly, variations on *e* are expected to be -8.66 × $10^{-6}$, -6.28 × $10^{-7}$, -2.15 × $10^{-7}$, -6.15 × $10^{-8}$, -9.61 × $10^{-9}$, -7.68 × $10^{-9}$.

Our numerical simulation offers a good match with analytical formulation (see the two Figures A5-A6 and Table A1). In particular, the use of a quite simplified force model greatly decreases numerical errors during integration making the comparison possible (up to two digits at least, which may be the accuracy of the analytical formulation we consider).

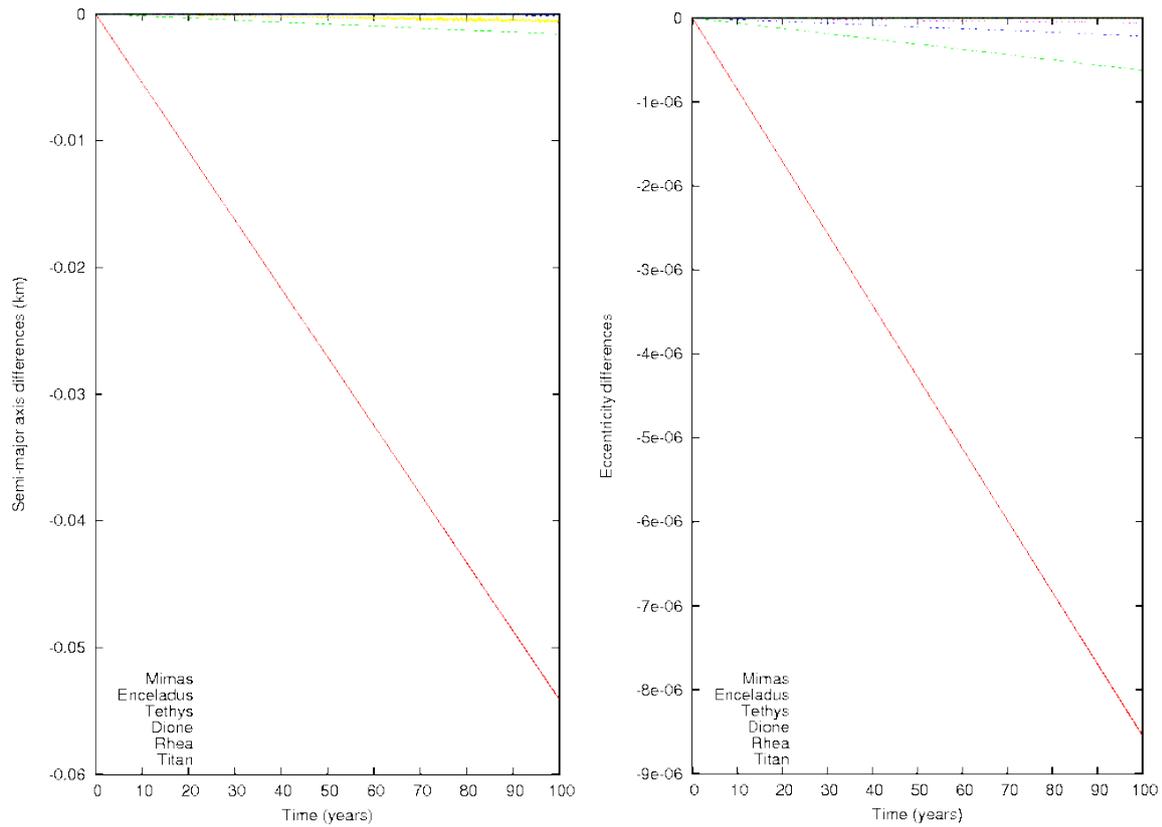

### 1.3.3 Tides in the planet (Full model):

One can try checking the tidal model with the full model considered in this work, even though perturbations will make the comparison with an analytical formulation more difficult. In particular, the expected drift on $a$ and $e$ may be masked by large short-period oscillations. Nevertheless, changes on $a$ can still be checked by looking at the associated acceleration in longitude. Expected variations over 100 years in $a$ from Eq. (A1) are 8.38, 6.08, 10.72, 4.89, 1.64, 0.95 meters for Mimas to Titan respectively. This translates to 1522, 761, 975, 306, 62, 10 km in the mean longitudes.

In the simulations below, Tethys and Dione have been removed from the model to avoid resonances (that would make the analytic formulation of Eq. (A1) invalid).

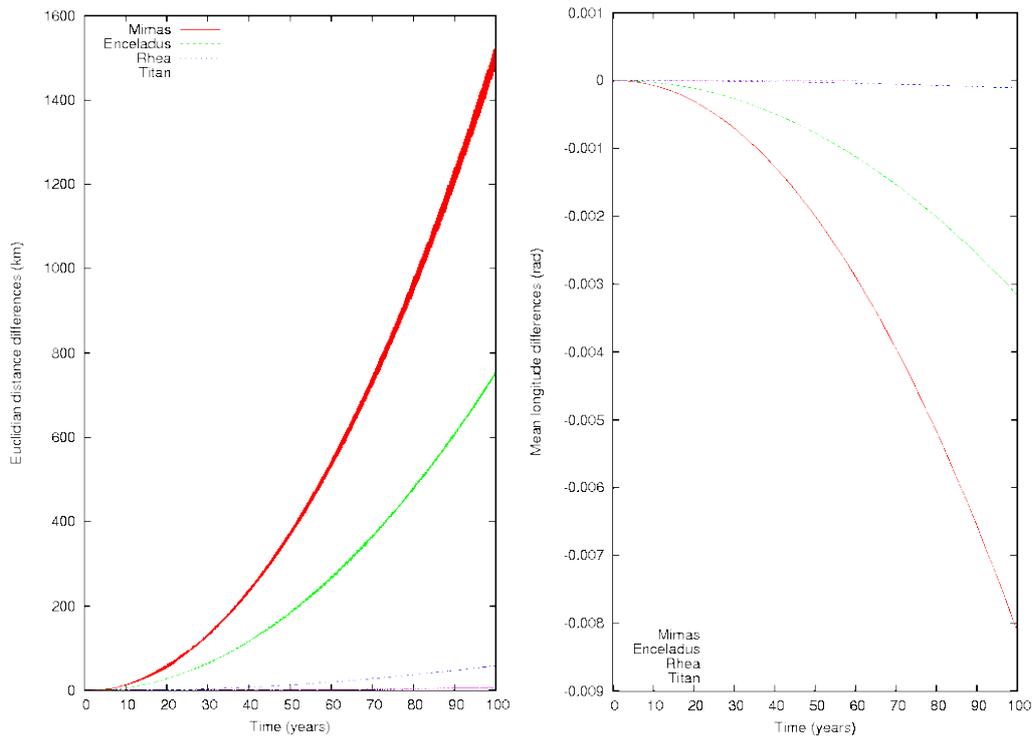

We obtain clearly a pretty good agreement (see Figures A7 and A8). A similar agreement can be found with the satellite case (Full model).

A.1.4 Testing da/dt acceleration:

As for the tidal model, the "da/dt" force cannot be tested by the conservation of energy. Using Gauss' equations, we recall that we have introduced a force on Mimas that affects only its semi-major axis by a constant drift da/dt. Checking the validity of this force can be done easily by checking Mimas' semi-major axis variations after integration.

Here we consider the difference between two numerical simulations, with/without Mimas' da/dt (Mimas' semi-major axis input) equal to $-16.9 \times 10^{-15}$ au/day.

A.1.4.1 Two-Body problem:

We provide below such a test (Figure A9), assuming a 2-body problem with only Saturn and Mimas. Clearly, the force acting on Mimas (input) and its expected effect on Mimas' semi-major axis (output) are in full agreement.

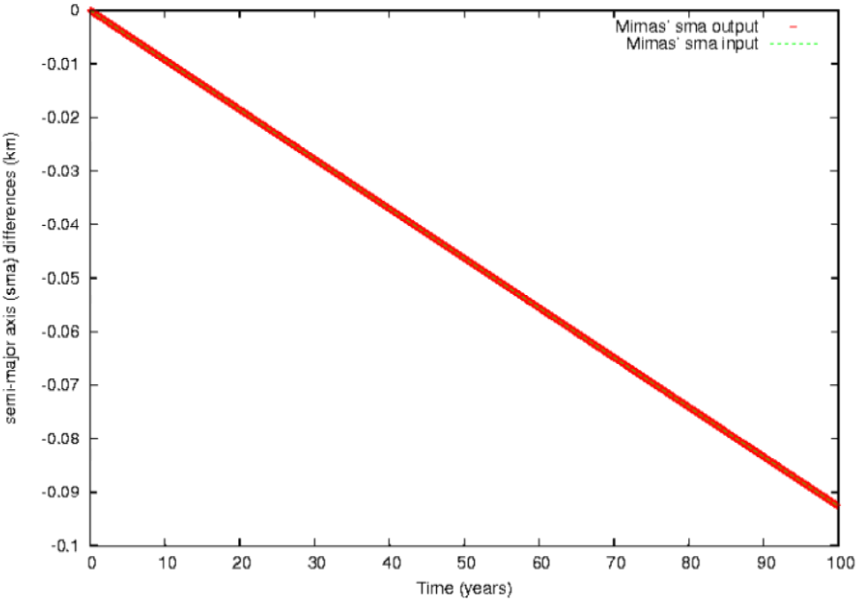

The other elements are not affected by the da/dt force, in the limit of accuracy of our integration.

A.1.4.2 Full model:

As for tidal effects, one may expect the comparison between input/output to be more difficult using the full model (Enceladus equilibrium scenario and constant Saturn's Q). Nevertheless, we can still use the mean longitude drift for the test and we have a pretty good agreement with the two Figures A10-A11 (Tethys has been removed from the simulation to avoid the Mimas-Tethys resonance). In particular, da/dt= -16.9 × $10^{-15}$ au/day translates to 16,776 km in Mimas' mean longitude after 100 years. The Figures A10-A11 show the variations in semi-major axis and in positions (which corresponds essentially to the mean longitude variations).

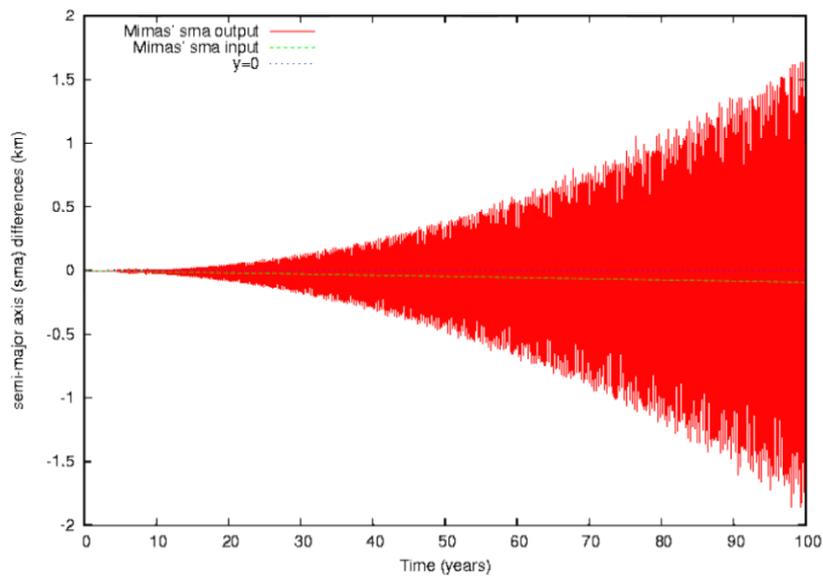

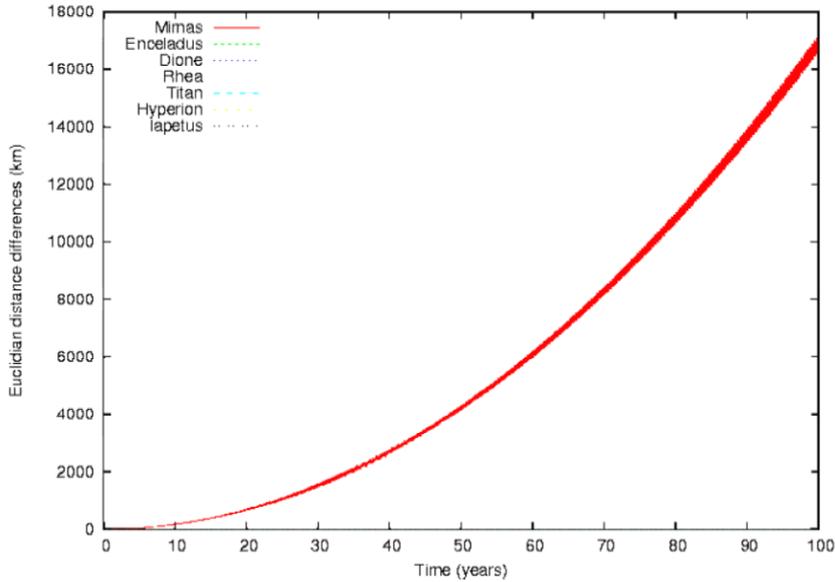

A.1.5 Testing the variational equations:

Partial derivatives of the moon state vectors as functions of initial conditions and physical parameters are computed by solving the so-called variational equations (see Eq. 4). Having accurate partial derivatives is a fundamental requirement when fitting a dynamical model to astrometric data. Solutions of the variational equations are routinely tested in our code. To check the accuracy of such computations, we compared our numerical solutions with their approximations using the centred difference method (i.e. $f'(x) \simeq [f(x+h) - f(x-h)]/2h$).

It is not possible to provide here the tests of all partial derivatives. Hence, we will restrict the number of figures by showing the computation of partial derivatives related to *da/dt* (Figures A12-A13), Saturn's *Q* (Figures A14-A15) and Enceladus' *Q* (denoted hereafter $Q^{s2}$; Figures A16-A17), only. Any other plots are available on request.

We recall that $c_l$ denotes an unspecified parameter of the model that shall be fitted. For more details on the method used, we refer to (Lainey et al. 2004).

Plots for $c_l = da/dt$ on Mimas position:

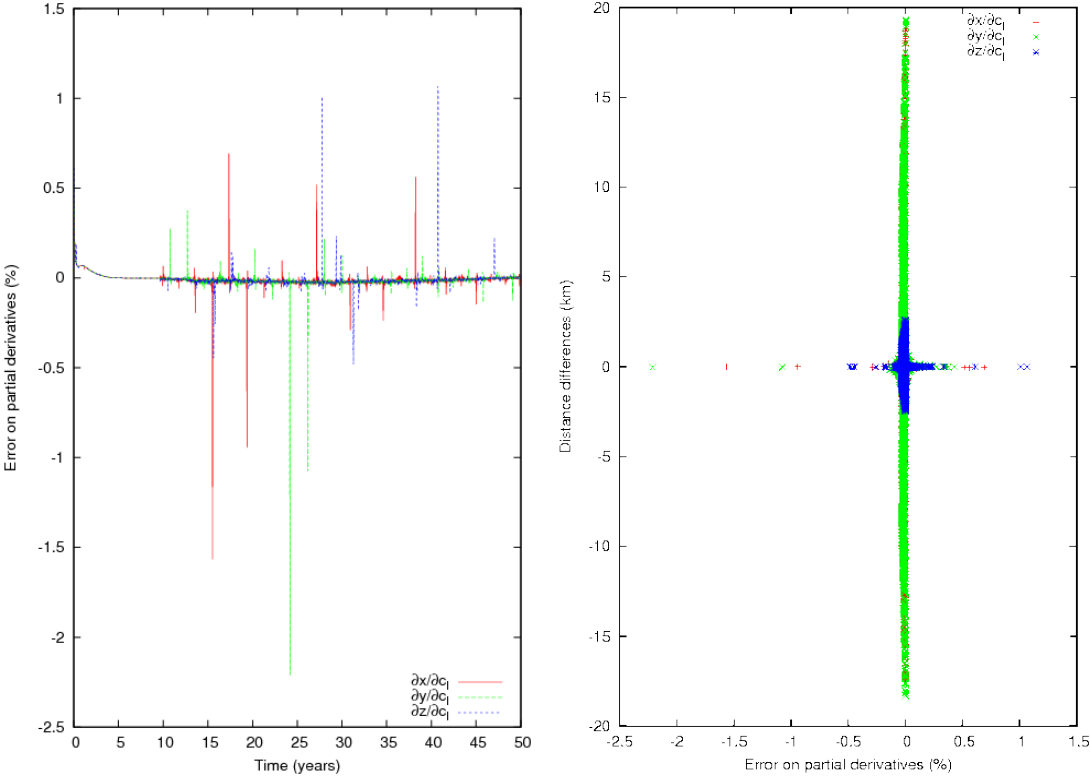

As can be seen in Figures A12-A13, numerical computation of $\partial r/\partial c_l$, where $r=r_1$ and $c_l=da/dt$, is in agreement with its approximation derived from the centre difference method. The scattering behaviour evident in Figure A12 is usual and corresponds to a non linear configuration occurring when the satellite reaches its maximum value along one Cartesian axis. As shown on the Figure A13, these configurations occur when the distance differences (on the considered axis) are pretty small. Hence, this does not affect the fitting procedure. See also (Lainey et al. 2004) for more details.

Plots for $c_l = Q$ on Mimas position (left) and Tethys position (right):

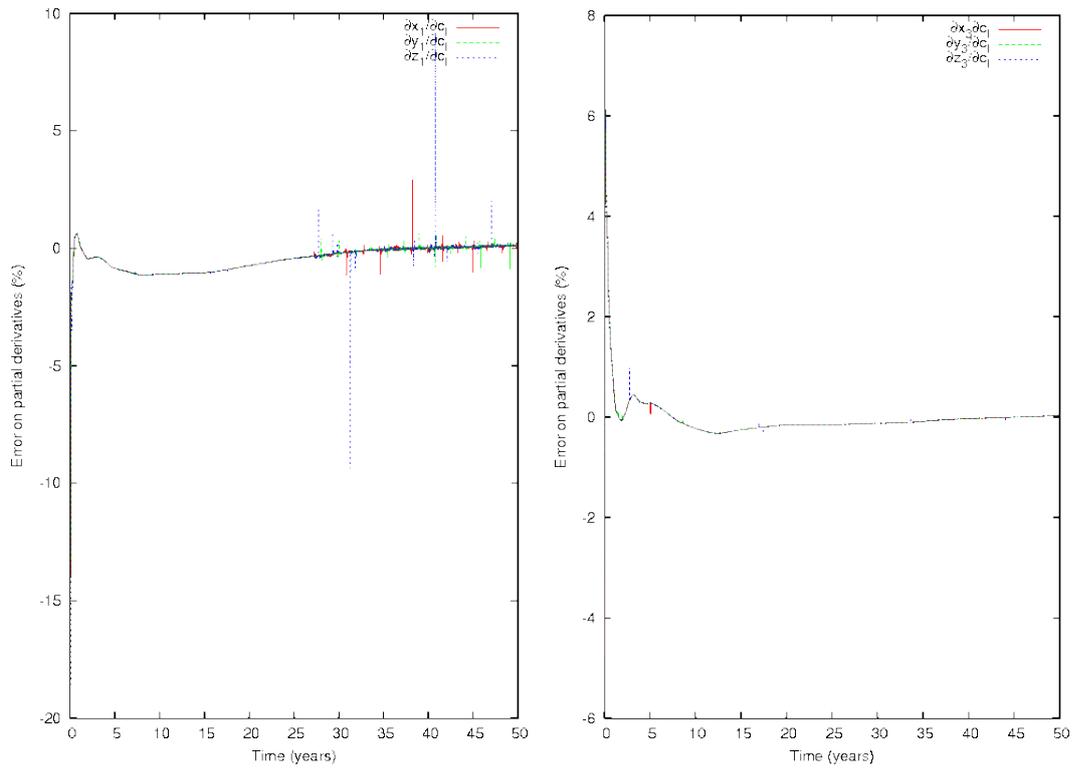

As can be seen on Figures A14-A15, numerical computation of $\partial r/\partial c_l$, where $\boldsymbol{r}=\boldsymbol{r_1}$ or $\boldsymbol{r}=\boldsymbol{r_3}$ and $c_l=Q$, is in agreement with its approximation derived from the centre difference method.

Plots for $c_l=Q^{s2}$ on Enceladus position (left) and Dione position (right) :

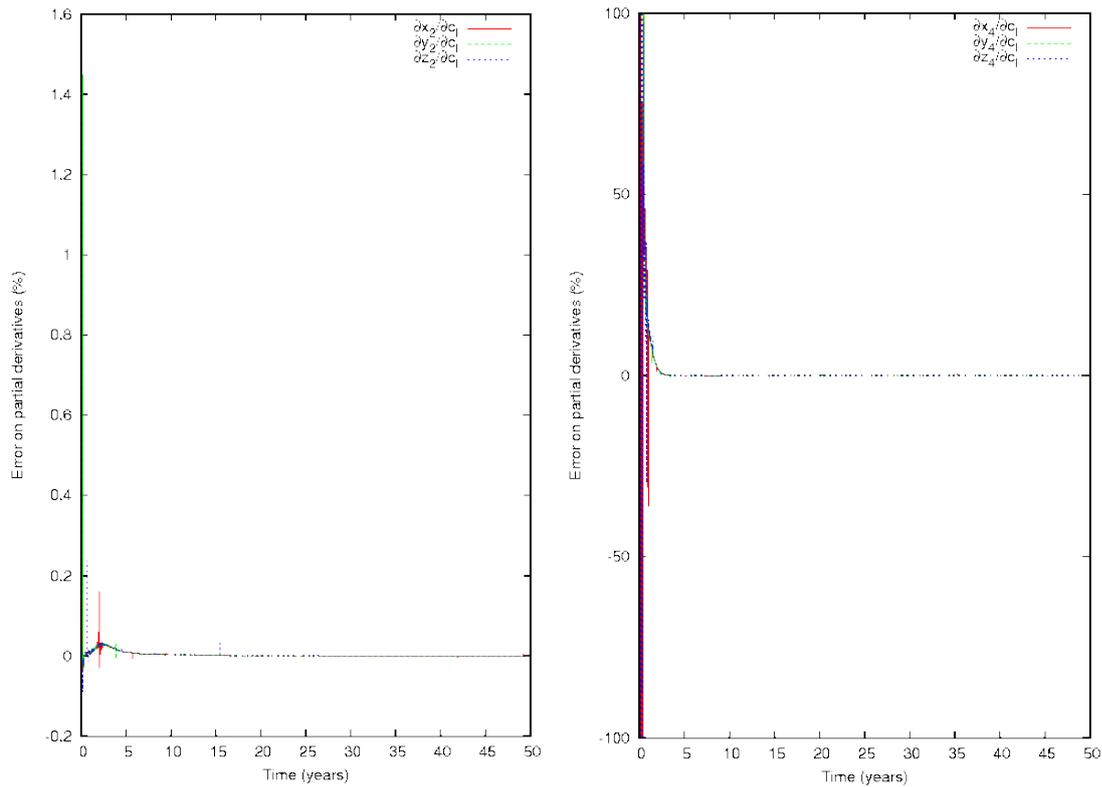

As can be seen on Figures A16-A17, numerical computation of $\partial r/\partial c_l$, where $r=r_2$ or $r=r_4$ and $c_l = Q^{s2}$, is in agreement with its approximation derived from the centre difference method.

### A.2 SPICE kernel NOE-6-2011-MAIN.bsp:

We have derived ephemerides of the eight main Saturnian moons for the Enceladus equilibrium and constant Q scenario. Our ephemerides are available as a SPICE kernel on the FTP server of IMCCE: ftp://ftp.imcce.fr/pub/ephem/satel/NOE/SATURNE/

The Figures A18-A19 show the Euclidian distance differences between our ephemerides and the JPL ones (kernel: sat317.bsp). Differences in the interval [1980,2011] are between a few tens of km to less than 200 km, except for Hyperion and Iapetus whose dynamics are less well-constrained by the astrometric observations we used.

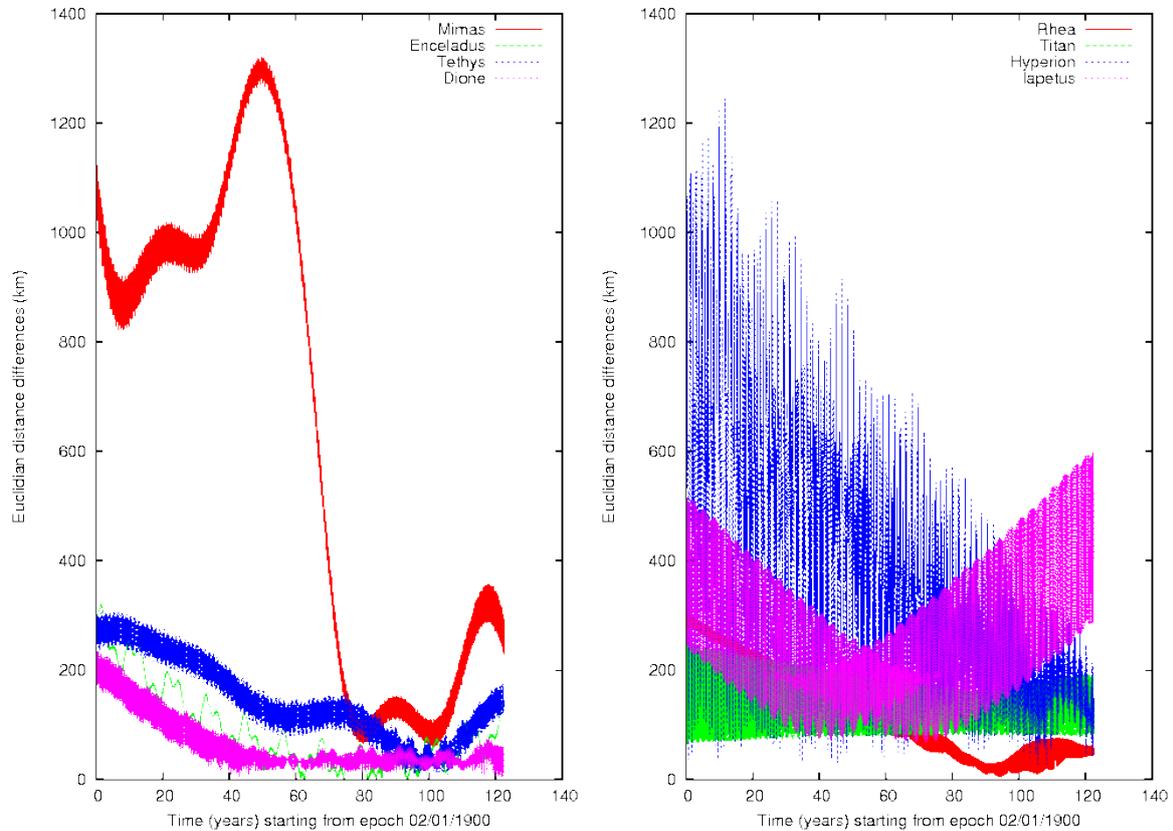

Since we have fitted a large *da/dt* term for Mimas, while JPL probably did not, one could expect large differences in Mimas' position. But as one can see, the large differences arise only before the 70s. This suggests that it is the use of a large time span which makes possible the derivation of *da/dt*.

A.3 Resonances:

While analytical developments require introducing explicitly orbital resonances, these latter are a simple consequences of initial conditions with numerical models. In the plots below, we provide the evolution of the resonant arguments associated to the Mimas-Tethys (left) and Enceladus-Dione resonances (right), respectively.

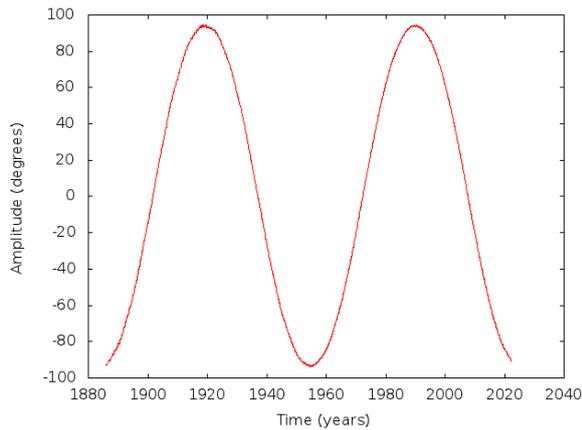 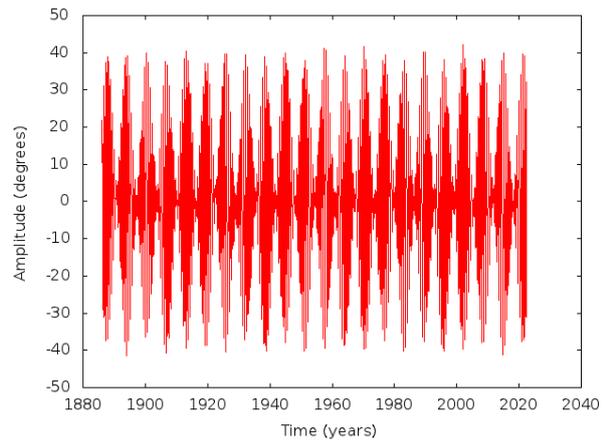

Figures:

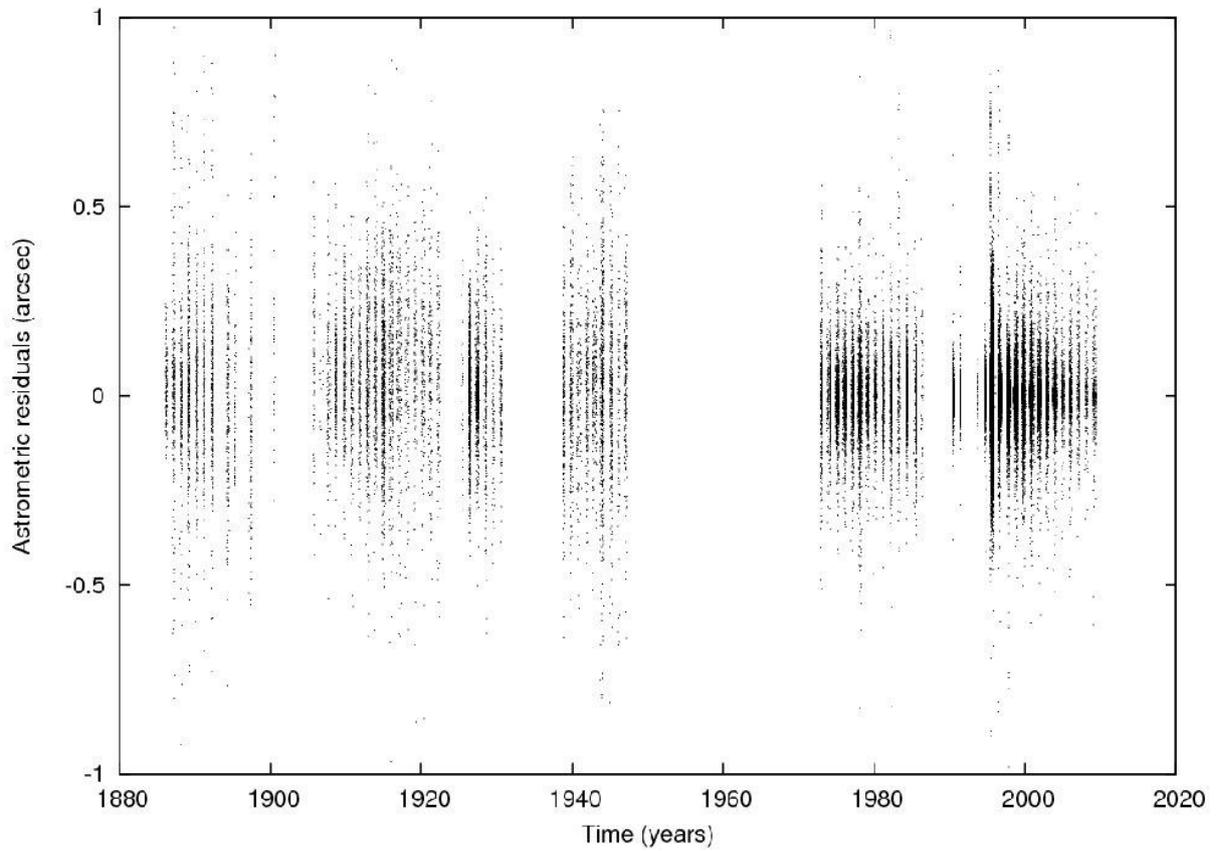

Fig. 1: **Astrometric residuals**

Residuals between the astrometric observations and our numerical model (assuming eccentricity equilibrium for Enceladus), after fitting the initial state vectors of the eight main Saturnian moons, the ratio $k_2/Q$ of Saturn and a constant rate *da/dt* on Mimas' semi-major axis. The global 1-$\sigma$ accuracy is about 0.1 arcsec (we recall that at the Saturnian distance 1 arcsec corresponds to about 6,000 km). It can be noted that no clear differences between old and modern observations are obvious due to: i) selective criteria in precision for all subsets (see Section 2 for details); ii) limitations inherent in graphical resolution (see Tables 1 and 2 for a detailed analysis of each observation subset).

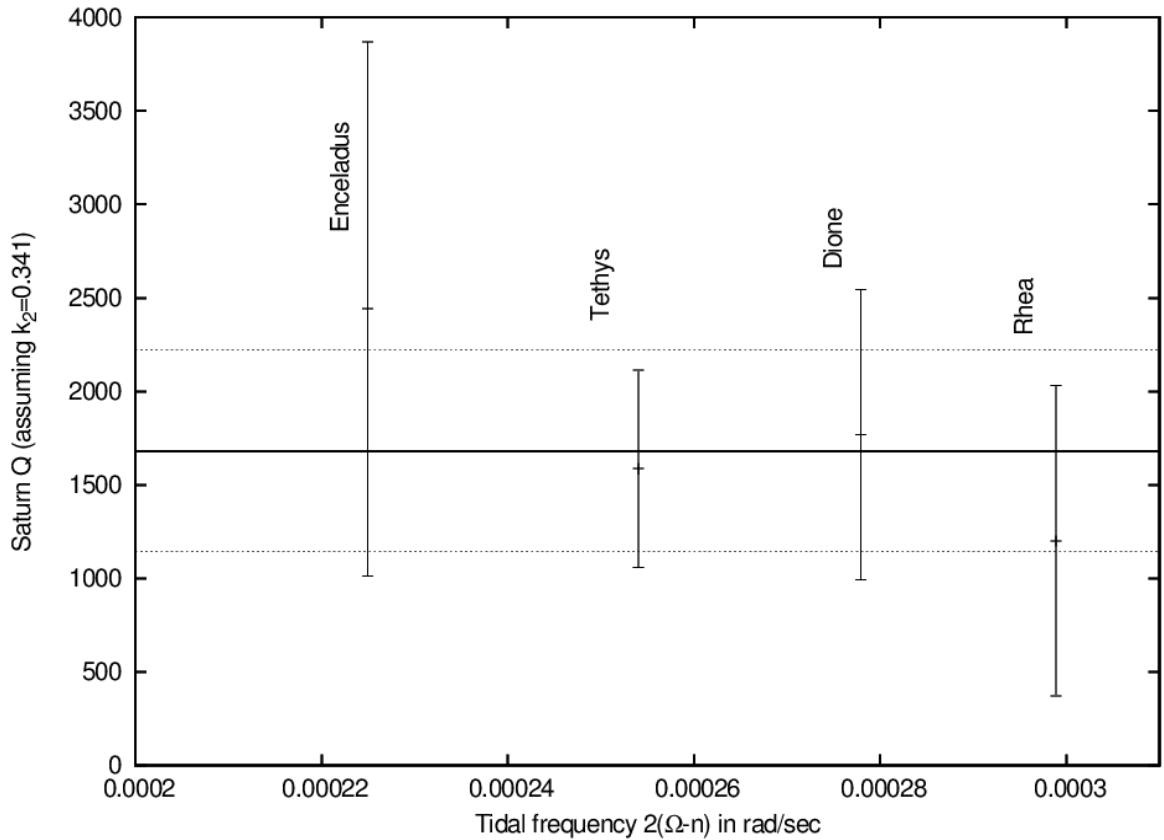

Fig. 2: **Determination of the Saturn tidal dissipation factor Q**

Saturn's tidal dissipation factor Q vs. the tidal frequency $2(\Omega-n)$, where $\Omega$ and n denote its rotation rate and the moon mean motion, respectively. The Love number is assumed to be $k_2=0.341$. The horizontal line shows the nominal solution (constant Q model) equal to $Q=1682 \pm 540$, with error bars as dashed lines. The vertical values with error bars are derived from a variable Q model. The large error bar associated with Enceladus is a consequence of the uncertain tidal dissipation in that satellite (whether one assumes no dissipation or eccentricity equilibrium). Presumably, the small value of Q is the signature of a rock-ice core or its boundary in Saturn.

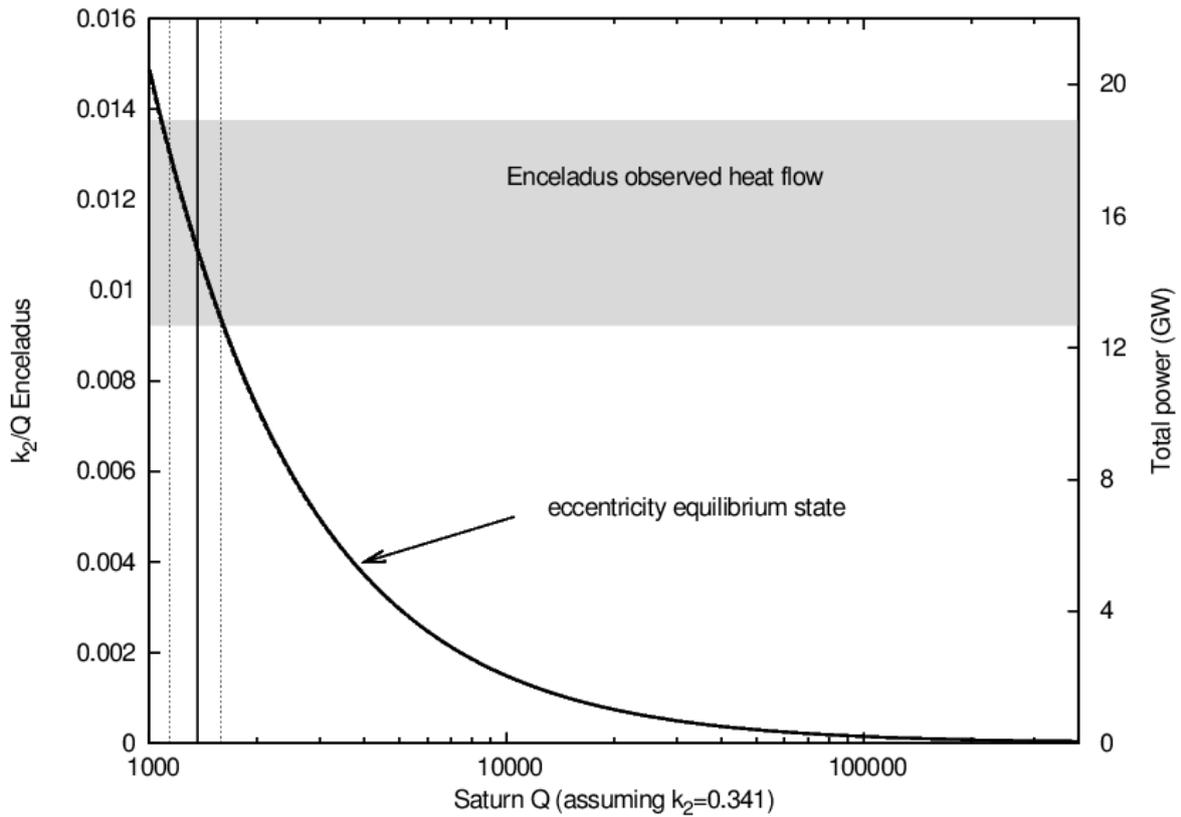

Fig. 3: **Comparison of Enceladus' thermal emission power with the Saturn tidal dissipation determined in the present study**

The solid curve indicates the $k_2/Q$ ratio and total dissipated power (in GW) in Enceladus for which the current orbital configuration of Enceladus and Dione is at eccentricity equilibrium as a function of Q in Saturn. Assuming such equilibrium in our fit, the value of Saturn's Q that we derive (Q = 1363 ± 221, see vertical line with associated dashed lines for error bars) shows that the expected total heat production rate is close to the observed emitted power. Hence, tidal heating equilibrium is a possible mechanism for maintaining Enceladus' thermal activity at its currently observed rate.

Tables:

Table 1- Statistics of the astrometric residuals computed from our model (Enceladus tidal equilibrium solution) in arcsecond. $\mu$ and $\sigma$ denote respectively the mean and standard deviation of the residuals computed in right ascension $\alpha.\cos(\delta)$ and declination $\delta$. $N_\alpha$ and $N_\delta$ are the number of observations considered for the respective coordinate. We recall that 0.1 second of arc corresponds to about 600 km at the Saturn distance.

| Observation subset: | $\nu_\alpha \cos(\delta)$ | $\sigma_{\alpha\cos(\delta)}$ | $\nu_\delta$ | $\sigma_\delta$ | $N_\alpha$, $N_\delta$ |
|---|---|---|---|---|---|
| All observations | | | | | |
| S1 | -0.0057 | 0.0952 | -0.0108 | 0.0725 | 371, 371 |
| S2 | 0.0019 | 0.1040 | 0.0028 | 0.1101 | 822, 822 |
| S3 | -0.0199 | 0.1267 | 0.0122 | 0.1067 | 1972, 1972 |
| S4 | 0.0020 | 0.1066 | 0.0113 | 0.1067 | 2271, 2271 |
| S5 | 0.0047 | 0.0899 | -0.0023 | 0.0863 | 2977, 2977 |
| S6 | 0.0121 | 0.1060 | -0.0171 | 0.1070 | 3271, 3271 |
| S7 | 0.1098 | 0.2984 | 0.0036 | 0.2166 | 973, 973 |
| S8 | 0.0140 | 0.1143 | -0.0052 | 0.1155 | 2008, 2008 |
| Alden H.L., O'Cornel W.C., (1928) | | | | | |
| S1 | 0.0000 | 0.0000 | 0.0000 | 0.0000 | 0, 0 |
| S2 | 0.0193 | 0.0890 | 0.0394 | 0.0816 | 40, 40 |
| S3 | 0.0267 | 0.0653 | 0.0066 | 0.0569 | 65, 65 |
| S4 | 0.0218 | 0.0493 | 0.0119 | 0.0467 | 64, 64 |
| S5 | 0.0007 | 0.0563 | 0.0204 | 0.0528 | 64, 64 |
| S6 | -0.0442 | 0.0681 | -0.0076 | 0.0566 | 64, 64 |
| S7 | 0.0000 | 0.0000 | 0.0000 | 0.0000 | 0, 0 |
| S8 | -0.0190 | 0.1538 | -0.0609 | 0.1337 | 59, 59 |
| Alden H.L. (1929) | | | | | |
| S1 | 0.0000 | 0.0000 | 0.0000 | 0.0000 | 0, 0 |
| S2 | 0.0258 | 0.0819 | 0.0228 | 0.0794 | 34, 34 |
| S3 | 0.0031 | 0.0420 | 0.0127 | 0.0526 | 38, 38 |
| S4 | 0.0097 | 0.0349 | 0.0359 | 0.0352 | 34, 34 |
| S5 | -0.0233 | 0.0520 | 0.0306 | 0.0364 | 36, 36 |
| S6 | -0.0267 | 0.0516 | -0.0132 | 0.0508 | 36, 36 |
| S7 | 0.0000 | 0.0000 | 0.0000 | 0.0000 | 0, 0 |
| S8 | 0.0135 | 0.0727 | -0.0887 | 0.1006 | 35, 35 |
| Sinclair A.T. (1974-1977) 13s | | | | | |
| S1 | 0.0000 | 0.0000 | 0.0000 | 0.0000 | 0, 0 |
| S2 | 0.0000 | 0.0000 | 0.0000 | 0.0000 | 0, 0 |
| S3 | 0.0279 | 0.1090 | -0.0107 | 0.1446 | 20, 20 |
| S4 | 0.0317 | 0.1551 | 0.0350 | 0.1528 | 25, 25 |
| S5 | 0.0040 | 0.1099 | -0.0330 | 0.0952 | 25, 25 |
| S6 | -0.1788 | 0.1799 | 0.0310 | 0.0819 | 25, 25 |
| S7 | 0.0000 | 0.0000 | 0.0000 | 0.0000 | 0, 0 |
| S8 | 0.1258 | 0.1603 | -0.0254 | 0.1186 | 24, 24 |
| Sinclair A.T. (1974-1977) 26s | | | | | |
| S1 | 0.0000 | 0.0000 | 0.0000 | 0.0000 | 0, 0 |
| S2 | 0.0000 | 0.0000 | 0.0000 | 0.0000 | 0, 0 |
| S3 | 0.0005 | 0.1069 | 0.0124 | 0.1011 | 40, 40 |
| S4 | 0.0044 | 0.0592 | 0.0098 | 0.0626 | 46, 46 |
| S5 | 0.0054 | 0.0698 | -0.0122 | 0.0805 | 48, 48 |
| S6 | -0.0002 | 0.0823 | -0.0079 | 0.0555 | 48, 48 |
| S7 | 0.0000 | 0.0000 | 0.0000 | 0.0000 | 0, 0 |
| S8 | -0.0099 | 0.0563 | 0.0004 | 0.0821 | 48, 48 |
| Abbot et al. (1975) PDS | | | | | |
| S1 | -0.1707 | 0.0000 | -0.0859 | 0.0000 | 1, 1 |
| S2 | -0.0472 | 0.0901 | 0.1828 | 0.1098 | 4, 4 |
| S3 | -0.0068 | 0.1408 | 0.1060 | 0.1260 | 10, 10 |
| S4 | -0.0476 | 0.1199 | 0.0084 | 0.0622 | 10, 10 |
| S5 | 0.0344 | 0.0347 | 0.0109 | 0.0371 | 10, 10 |
| S6 | -0.0693 | 0.0702 | -0.0587 | 0.0241 | 11, 11 |
| S7 | -0.2607 | 0.1166 | 0.3553 | 0.0757 | 6, 6 |
| S8 | 0.0138 | 0.0739 | 0.0475 | 0.0274 | 10, 10 |

| Observation subset: | $\nu_\alpha \cos(\delta)$ | $\sigma_{\alpha\cos(\delta)}$ | $\nu_\delta$ | $\sigma_\delta$ | $N_\alpha$, $N_\delta$ |
|---|---|---|---|---|---|
| Abbot et al. (1975) Mann | | | | | |
| S1 | 0.0692 | 0.0000 | -0.1231 | 0.0000 | 1, 1 |
| S2 | -0.1171 | 0.0981 | 0.3500 | 0.2047 | 5, 5 |
| S3 | 0.0028 | 0.0888 | 0.0298 | 0.1325 | 11, 11 |
| S4 | -0.0321 | 0.1280 | 0.0584 | 0.1849 | 11, 11 |
| S5 | 0.0393 | 0.0448 | 0.0136 | 0.0761 | 11, 11 |
| S6 | -0.0960 | 0.1324 | -0.0234 | 0.0385 | 11, 11 |
| S7 | -0.2767 | 0.0599 | 0.4871 | 0.0485 | 6, 6 |
| S8 | 0.0325 | 0.0941 | 0.0465 | 0.0559 | 11, 11 |
| Voronenko et al. (1991) | | | | | |
| S1 | 0.0000 | 0.0000 | 0.0000 | 0.0000 | 0, 0 |
| S2 | 0.0000 | 0.0000 | 0.0000 | 0.0000 | 0, 0 |
| S3 | 0.0172 | 0.2064 | -0.0056 | 0.1350 | 85, 85 |
| S4 | -0.0324 | 0.1675 | 0.0066 | 0.1712 | 96, 96 |
| S5 | -0.0034 | 0.1455 | 0.0214 | 0.0976 | 143, 143 |
| S6 | 0.0179 | 0.1319 | -0.0196 | 0.1084 | 153, 153 |
| S7 | 0.0000 | 0.0000 | 0.0000 | 0.0000 | 0, 0 |
| S8 | 0.0154 | 0.1679 | -0.0263 | 0.1085 | 19, 19 |
| Pascu (1982) | | | | | |
| S1 | -0.0105 | 0.1620 | -0.0393 | 0.1155 | 56, 56 |
| S2 | 0.0041 | 0.0953 | -0.0228 | 0.1208 | 107, 107 |
| S3 | 0.0148 | 0.0644 | -0.0104 | 0.0775 | 138, 138 |
| S4 | -0.0001 | 0.0505 | -0.0064 | 0.0608 | 163, 163 |
| S5 | 0.0115 | 0.0502 | -0.0041 | 0.0602 | 209, 209 |
| S6 | -0.0010 | 0.0529 | 0.0153 | 0.0644 | 228, 228 |
| S7 | 0.0761 | 0.2466 | -0.0564 | 0.1824 | 11, 11 |
| S8 | -0.0191 | 0.0886 | 0.0211 | 0.1071 | 213, 213 |
| Tolbin S.B. (1991) | | | | | |
| S1 | -0.0297 | 0.1602 | -0.0414 | 0.1813 | 21, 21 |
| S2 | 0.0059 | 0.0884 | 0.0090 | 0.1327 | 57, 57 |
| S3 | 0.0004 | 0.0697 | -0.0180 | 0.0834 | 75, 75 |
| S4 | 0.0127 | 0.0551 | -0.0073 | 0.0856 | 81, 81 |
| S5 | 0.0016 | 0.0526 | 0.0047 | 0.0684 | 88, 88 |
| S6 | 0.0033 | 0.0758 | -0.0022 | 0.0968 | 89, 89 |
| S7 | 0.0000 | 0.0000 | 0.0000 | 0.0000 | 0, 0 |
| S8 | -0.0194 | 0.1521 | 0.0336 | 0.1261 | 62, 62 |
| Tolbin S.B. (1991) | | | | | |
| S1 | 0.0292 | 0.1520 | 0.0194 | 0.1375 | 7, 7 |
| S2 | -0.0120 | 0.1237 | 0.0266 | 0.1293 | 50, 50 |
| S3 | 0.0003 | 0.0596 | -0.0037 | 0.0767 | 89, 89 |
| S4 | 0.0121 | 0.0587 | -0.0006 | 0.0706 | 96, 96 |
| S5 | 0.0012 | 0.0617 | -0.0041 | 0.0817 | 102, 102 |
| S6 | 0.0067 | 0.0674 | -0.0354 | 0.1222 | 107, 107 |
| S7 | 0.0000 | 0.0000 | 0.0000 | 0.0000 | 0, 0 |
| S8 | -0.0189 | 0.0740 | 0.0309 | 0.1920 | 80, 80 |
| Seitzer & Ianna (1980) | | | | | |
| S1 | 0.0000 | 0.0000 | 0.0000 | 0.0000 | 0, 0 |
| S2 | 0.0000 | 0.0000 | 0.0000 | 0.0000 | 0, 0 |
| S3 | -0.0879 | 0.0096 | 0.0504 | 0.0297 | 3, 3 |
| S4 | -0.0332 | 0.0198 | -0.0141 | 0.0109 | 3, 3 |
| S5 | 0.0654 | 0.0629 | -0.1057 | 0.0739 | 10, 10 |
| S6 | 0.0454 | 0.0849 | -0.0381 | 0.0600 | 17, 17 |
| S7 | 0.0000 | 0.0000 | 0.0000 | 0.0000 | 0, 0 |
| S8 | -0.0515 | 0.1236 | 0.0060 | 0.1843 | 24, 24 |

| Observation subset: | $\nu_\alpha \cos(\delta)$ | $\sigma_{\alpha\cos(\delta)}$ | $\nu_\delta$ | $\sigma_\delta$ | $N_\alpha$, $N_\delta$ |
|---|---|---|---|---|---|
| Taylor & Sinclair (1985) | | | | | |
| S1 | 0.0000 | 0.0000 | 0.0000 | 0.0000 | 0, 0 |
| S2 | 0.0604 | 0.1549 | 0.0443 | 0.0569 | 10, 10 |
| S3 | 0.0710 | 0.1993 | 0.0589 | 0.0852 | 20, 20 |
| S4 | -0.0485 | 0.1979 | 0.0097 | 0.1309 | 35, 35 |
| S5 | 0.0250 | 0.1227 | -0.0005 | 0.1180 | 38, 38 |
| S6 | -0.0078 | 0.0973 | -0.0696 | 0.1138 | 45, 45 |
| S7 | 0.2277 | 0.5327 | 0.0817 | 0.5884 | 38, 38 |
| S8 | -0.0152 | 0.1239 | 0.0405 | 0.1432 | 45, 45 |
| Seitzer et al. (1979) | | | | | |
| S1 | -0.1067 | 0.1613 | -0.0564 | 0.0106 | 2, 2 |
| S2 | 0.0000 | 0.0000 | 0.0000 | 0.0000 | 0, 0 |
| S3 | 0.0055 | 0.0678 | -0.0073 | 0.0849 | 49, 49 |
| S4 | 0.0539 | 0.2094 | -0.1229 | 0.1929 | 41, 41 |
| S5 | 0.0082 | 0.0652 | -0.0387 | 0.0756 | 49, 49 |
| S6 | 0.0829 | 0.1042 | -0.0467 | 0.1039 | 50, 50 |
| S7 | -0.3379 | 0.1333 | 0.0453 | 0.1124 | 4, 4 |
| S8 | -0.0430 | 0.0971 | 0.1102 | 0.1839 | 60, 60 |
| Dourneau et al. (1986) | | | | | |
| S1 | 0.0686 | 0.2278 | -0.0274 | 0.1042 | 11, 11 |
| S2 | 0.0146 | 0.1244 | -0.0151 | 0.0593 | 39, 39 |
| S3 | 0.0249 | 0.0786 | -0.0285 | 0.0488 | 39, 39 |
| S4 | -0.0259 | 0.0691 | -0.0260 | 0.0570 | 56, 56 |
| S5 | 0.0045 | 0.0644 | -0.0097 | 0.0478 | 76, 76 |
| S6 | 0.0101 | 0.0684 | 0.0549 | 0.0426 | 82, 82 |
| S7 | 0.7665 | 0.1130 | 0.0276 | 0.0832 | 95, 95 |
| S8 | -0.0126 | 0.1165 | -0.0019 | 0.0698 | 95, 95 |
| Dourneau & Veillet 3.6 | | | | | |
| S1 | 0.0000 | 0.0000 | 0.0000 | 0.0000 | 0, 0 |
| S2 | 0.0000 | 0.0000 | 0.0000 | 0.0000 | 0, 0 |
| S3 | 0.0037 | 0.0858 | -0.0083 | 0.1542 | 20, 20 |
| S4 | 0.0630 | 0.1749 | -0.0281 | 0.1058 | 25, 25 |
| S5 | 0.0651 | 0.0404 | -0.0392 | 0.0561 | 14, 14 |
| S6 | -0.0164 | 0.1288 | -0.1052 | 0.0773 | 17, 17 |
| S7 | 0.1865 | 0.3238 | -0.0286 | 0.0985 | 25, 25 |
| S8 | -0.0763 | 0.1239 | 0.1069 | 0.0649 | 30, 30 |
| Dourneau & Veillet 1.5 | | | | | |
| S1 | -0.1346 | 0.1196 | -0.0500 | 0.0749 | 10, 10 |
| S2 | -0.0014 | 0.1013 | -0.0223 | 0.0880 | 57, 57 |
| S3 | -0.0083 | 0.1103 | 0.0007 | 0.0807 | 78, 78 |
| S4 | 0.0112 | 0.0689 | -0.0032 | 0.0674 | 155, 155 |
| S5 | 0.0028 | 0.0542 | -0.0019 | 0.0495 | 195, 195 |
| S6 | -0.0015 | 0.0671 | 0.0167 | 0.0479 | 197, 197 |
| S7 | 0.0006 | 0.1234 | 0.0118 | 0.0916 | 197, 197 |
| S8 | 0.0004 | 0.0874 | -0.0037 | 0.0976 | 196, 196 |
| Kiseleva et al. (1996) | | | | | |
| S1 | 0.0000 | 0.0000 | 0.0000 | 0.0000 | 0, 0 |
| S2 | -0.0420 | 0.0940 | 0.0219 | 0.1574 | 10, 10 |
| S3 | -0.0168 | 0.0868 | 0.0413 | 0.1150 | 11, 11 |
| S4 | -0.0057 | 0.0507 | -0.0251 | 0.0566 | 25, 25 |
| S5 | -0.0211 | 0.0623 | -0.0188 | 0.0792 | 25, 25 |
| S6 | -0.0036 | 0.0592 | -0.0081 | 0.0834 | 32, 32 |
| S7 | 0.0000 | 0.0000 | 0.0000 | 0.0000 | 0, 0 |
| S8 | 0.0662 | 0.0902 | 0.0325 | 0.1311 | 21, 21 |

| Observation subset: | $\nu_\alpha \cos(\delta)$ | $\sigma_{\alpha\cos(\delta)}$ | $\nu_\delta$ | $\sigma_\delta$ | $N_\alpha, N_\delta$ |
|---|---|---|---|---|---|
| Vass-G. (1997) | | | | | |
| S1 | 0.0000 | 0.0000 | 0.0000 | 0.0000 | 0, 0 |
| S2 | 0.0025 | 0.1647 | -0.0132 | 0.1532 | 151, 151 |
| S3 | -0.0704 | 0.1494 | 0.0319 | 0.1246 | 654, 654 |
| S4 | -0.0039 | 0.1388 | 0.0480 | 0.1269 | 600, 600 |
| S5 | 0.0061 | 0.1237 | -0.0059 | 0.1135 | 948, 948 |
| S6 | 0.0083 | 0.1163 | -0.0203 | 0.1193 | 1287, 1287 |
| S7 | -0.0477 | 0.1564 | 0.0478 | 0.2366 | 94, 94 |
| S8 | 0.1306 | 0.1505 | -0.0654 | 0.1037 | 243, 243 |
| Kisseleva & Kalin.(2000) | | | | | |
| S1 | -0.1235 | 0.0434 | 0.0449 | 0.0373 | 3, 3 |
| S2 | -0.0387 | 0.0923 | 0.0023 | 0.0716 | 10, 10 |
| S3 | -0.0165 | 0.0654 | 0.0182 | 0.0748 | 22, 22 |
| S4 | 0.0239 | 0.0527 | -0.0236 | 0.0686 | 23, 23 |
| S5 | 0.0039 | 0.0532 | 0.0138 | 0.0594 | 27, 27 |
| S6 | 0.0100 | 0.0564 | 0.0069 | 0.0768 | 27, 27 |
| S7 | 0.0000 | 0.0000 | 0.0000 | 0.0000 | 0, 0 |
| S8 | 0.0139 | 0.1079 | -0.0410 | 0.1718 | 14, 14 |
| Kiseleva et al. (1998) | | | | | |
| S1 | 0.0000 | 0.0000 | 0.0000 | 0.0000 | 0, 0 |
| S2 | -0.0768 | 0.1335 | 0.2952 | 0.1175 | 3, 3 |
| S3 | -0.0178 | 0.0415 | -0.0342 | 0.1405 | 7, 7 |
| S4 | 0.0509 | 0.0714 | -0.0259 | 0.1535 | 4, 4 |
| S5 | -0.0124 | 0.0384 | -0.0070 | 0.0762 | 10, 10 |
| S6 | 0.0386 | 0.0718 | -0.0818 | 0.2194 | 12, 12 |
| S7 | 0.0000 | 0.0000 | 0.0000 | 0.0000 | 0, 0 |
| S8 | 0.0117 | 0.0000 | -0.0871 | 0.0000 | 1, 1 |
| French(2006) HST-WF4 | | | | | |
| S1 | -0.0002 | 0.0111 | -0.0093 | 0.0104 | 39, 39 |
| S2 | 0.0025 | 0.0132 | 0.0005 | 0.0157 | 53, 53 |
| S3 | 0.0039 | 0.0167 | 0.0057 | 0.0160 | 63, 63 |
| S4 | 0.0027 | 0.0196 | 0.0054 | 0.0269 | 33, 33 |
| S5 | -0.0011 | 0.0152 | 0.0014 | 0.0199 | 39, 39 |
| S6 | -0.0273 | 0.0401 | -0.0107 | 0.0328 | 23, 23 |
| S7 | 0.0856 | 0.0074 | 0.0009 | 0.0040 | 4, 4 |
| S8 | 0.0000 | 0.0000 | 0.0000 | 0.0000 | 0, 0 |
| French(2006) HST-PC | | | | | |
| S1 | 0.0022 | 0.0076 | 0.0004 | 0.0098 | 154, 154 |
| S2 | -0.0027 | 0.0056 | -0.0001 | 0.0087 | 82, 82 |
| S3 | 0.0056 | 0.0124 | 0.0031 | 0.0077 | 24, 24 |
| S4 | -0.0112 | 0.0024 | -0.0122 | 0.0013 | 5, 5 |
| S5 | 0.0083 | 0.0060 | 0.0049 | 0.0081 | 10, 10 |
| S6 | 0.0000 | 0.0000 | 0.0000 | 0.0000 | 0, 0 |
| S7 | 0.0000 | 0.0000 | 0.0000 | 0.0000 | 0, 0 |
| S8 | 0.0000 | 0.0000 | 0.0000 | 0.0000 | 0, 0 |
| French(2006) HST-WF3 | | | | | |
| S1 | 0.0037 | 0.0079 | -0.0030 | 0.0173 | 25, 25 |
| S2 | -0.0034 | 0.0117 | -0.0026 | 0.0130 | 51, 51 |
| S3 | 0.0011 | 0.0086 | 0.0007 | 0.0142 | 55, 55 |
| S4 | 0.0024 | 0.0187 | -0.0087 | 0.0193 | 99, 99 |
| S5 | -0.0022 | 0.0166 | 0.0012 | 0.0291 | 70, 70 |
| S6 | 0.0000 | 0.0000 | 0.0000 | 0.0000 | 0, 0 |
| S7 | -0.0119 | 0.0095 | -0.0148 | 0.0301 | 13, 13 |
| S8 | 0.0000 | 0.0000 | 0.0000 | 0.0000 | 0, 0 |

| Observation subset: | $\nu_\alpha \cos(\delta)$ | $\sigma_{\alpha\cos(\delta)}$ | $\nu_\delta$ | $\sigma_\delta$ | $N_\alpha, N_\delta$ |
|---|---|---|---|---|---|
| French(2006) HST-WF2 | | | | | |
| S1 | 0.0022 | 0.0119 | 0.0034 | 0.0135 | 34, 34 |
| S2 | -0.0020 | 0.0160 | 0.0119 | 0.0111 | 35, 35 |
| S3 | 0.0042 | 0.0111 | -0.0073 | 0.0206 | 33, 33 |
| S4 | 0.0089 | 0.0199 | 0.0074 | 0.0281 | 87, 87 |
| S5 | -0.0127 | 0.0140 | -0.0105 | 0.0223 | 35, 35 |
| S6 | 0.0000 | 0.0000 | 0.0000 | 0.0000 | 0, 0 |
| S7 | 0.0030 | 0.0509 | 0.0234 | 0.0189 | 30, 30 |
| S8 | 0.0000 | 0.0000 | 0.0000 | 0.0000 | 0, 0 |
| Flagstaff | | | | | |
| S1 | 0.0000 | 0.0000 | 0.0000 | 0.0000 | 0, 0 |
| S2 | 0.0000 | 0.0000 | 0.0000 | 0.0000 | 0, 0 |
| S3 | -0.0034 | 0.1461 | 0.0164 | 0.1315 | 251, 251 |
| S4 | 0.0059 | 0.0912 | 0.0089 | 0.1031 | 398, 398 |
| S5 | 0.0040 | 0.0585 | 0.0008 | 0.0721 | 651, 651 |
| S6 | 0.0458 | 0.1052 | -0.0335 | 0.1124 | 682, 682 |
| S7 | 0.0635 | 0.1851 | -0.0299 | 0.2159 | 450, 450 |
| S8 | 0.0094 | 0.0752 | -0.0144 | 0.0801 | 705, 705 |
| Kiseleva (unpublished) | | | | | |
| S1 | -0.0225 | 0.0601 | 0.0618 | 0.0327 | 3, 3 |
| S2 | 0.0428 | 0.0790 | 0.0009 | 0.0523 | 10, 10 |
| S3 | 0.0061 | 0.0896 | -0.0300 | 0.1058 | 19, 19 |
| S4 | -0.0152 | 0.0794 | -0.0495 | 0.1321 | 20, 20 |
| S5 | 0.0107 | 0.0581 | -0.0109 | 0.0700 | 21, 21 |
| S6 | -0.0208 | 0.0822 | -0.0323 | 0.0945 | 25, 25 |
| S7 | 0.0000 | 0.0000 | 0.0000 | 0.0000 | 0, 0 |
| S8 | 0.0093 | 0.0903 | 0.1847 | 0.1637 | 13, 13 |
| PHESAT | | | | | |
| S1 | -0.0653 | 0.0649 | -0.0073 | 0.0099 | 4, 4 |
| S2 | 0.0009 | 0.0203 | 0.0046 | 0.0103 | 14, 14 |
| S3 | -0.0021 | 0.0152 | 0.0027 | 0.0180 | 53, 53 |
| S4 | 0.0002 | 0.0281 | 0.0024 | 0.0180 | 34, 34 |
| S5 | 0.0072 | 0.0179 | -0.0005 | 0.0289 | 23, 23 |
| S6 | 0.0010 | 0.0018 | 0.0100 | 0.0272 | 3, 3 |
| S7 | 0.0000 | 0.0000 | 0.0000 | 0.0000 | 0, 0 |
| S8 | 0.0000 | 0.0000 | 0.0000 | 0.0000 | 0, 0 |

Table 2- Statistics of the astrometric residuals computed from our model (Enceladus tidal equilibrium solution) in arcsecond. $\mu$ and $\sigma$ denote respectively the mean and standard deviation of the residuals computed in separation $s$ and position angle $p$. $N_s$ *and* $N_p$ are the number of observations considered for the respective coordinate. We recall that 0.1 second of arc corresponds to about 600 km at the Saturn distance.

| Observation subset: | $\nu_s$ | $\sigma_s$ | $\nu_p$ | $\sigma_p$ | $N_s$, $N_p$ |
|---|---|---|---|---|---|
| All Observations: | | | | | |
| S1 | 0.0140 | 0.1027 | 0.0131 | 0.1152 | 1285, 1298 |
| S2 | -0.0032 | 0.0988 | 0.0048 | 0.1069 | 2640, 2643 |
| S3 | 0.0157 | 0.1130 | -0.0003 | 0.1152 | 4702, 4700 |
| S4 | 0.0150 | 0.1045 | 0.0023 | 0.1096 | 3775, 3776 |
| S5 | 0.0113 | 0.1088 | 0.0030 | 0.1151 | 4471, 4489 |
| S6 | 0.0238 | 0.0937 | -0.0049 | 0.1084 | 2842, 2836 |
| S7 | 0.0017 | 0.3275 | 0.1068 | 0.4838 | 138, 113 |
| S8 | 0.0179 | 0.0766 | 0.0076 | 0.1246 | 1098, 1101 |
| Struve(1898) 61/62 | | | | | |
| S1 | 0.0076 | 0.1756 | 0.0394 | 0.1810 | 105, 119 |
| S2 | -0.0011 | 0.1129 | -0.0215 | 0.1170 | 218, 226 |
| S3 | 0.0617 | 0.1363 | -0.0185 | 0.1190 | 276, 281 |
| S4 | 0.0671 | 0.1367 | -0.0242 | 0.1252 | 167, 170 |
| S5 | 0.0000 | 0.0000 | 0.0000 | 0.0000 | 0, 0 |
| S6 | 0.0000 | 0.0000 | 0.0000 | 0.0000 | 0, 0 |
| S7 | 0.0000 | 0.0000 | 0.0000 | 0.0000 | 0, 0 |
| S8 | 0.0531 | 0.0562 | -0.0140 | 0.1181 | 6, 6 |
| Struve(1898) 21/22 | | | | | |
| S1 | 0.0000 | 0.0000 | 0.0000 | 0.0000 | 0, 0 |
| S2 | 0.0000 | 0.0000 | 0.0000 | 0.0000 | 0, 0 |
| S3 | 0.0000 | 0.0000 | 0.0000 | 0.0000 | 0, 0 |
| S4 | 0.0000 | 0.0000 | 0.0000 | 0.0000 | 0, 0 |
| S5 | 0.0000 | 0.0000 | 0.0000 | 0.0000 | 0, 0 |
| S6 | 0.0000 | 0.0000 | 0.0000 | 0.0000 | 0, 0 |
| S7 | 0.0017 | 0.3275 | 0.1068 | 0.4838 | 138, 113 |
| S8 | -0.0930 | 0.2411 | -0.0063 | 0.5614 | 4, 4 |
| Struve(1898) | | | | | |
| S1 | 0.0000 | 0.0000 | 0.0000 | 0.0000 | 0, 0 |
| S2 | 0.0000 | 0.0000 | 0.0000 | 0.0000 | 0, 0 |
| S3 | 0.0000 | 0.0000 | 0.0000 | 0.0000 | 0, 0 |
| S4 | 0.0000 | 0.0000 | 0.0000 | 0.0000 | 0, 0 |
| S5 | -0.1310 | 0.1191 | 0.1167 | 0.0953 | 44, 42 |
| S6 | 0.0339 | 0.1579 | 0.1721 | 0.1457 | 57, 54 |
| S7 | 0.0000 | 0.0000 | 0.0000 | 0.0000 | 0, 0 |
| S8 | 0.0000 | 0.0000 | 0.0000 | 0.0000 | 0, 0 |
| Stone (1895b) | | | | | |
| S1 | 0.0000 | 0.0000 | 0.0000 | 0.0000 | 0, 0 |
| S2 | -0.0725 | 0.1892 | -0.0533 | 0.1717 | 5, 5 |
| S3 | -0.0654 | 0.1987 | -0.0908 | 0.1791 | 16, 18 |
| S4 | -0.0730 | 0.1514 | 0.0623 | 0.1837 | 17, 19 |
| S5 | 0.0051 | 0.0761 | -0.0449 | 0.2588 | 6, 6 |
| S6 | 0.0750 | 0.0000 | 0.0126 | 0.0998 | 1, 2 |
| S7 | 0.0000 | 0.0000 | 0.0000 | 0.0000 | 0, 0 |
| S8 | 0.0000 | 0.0000 | 0.0000 | 0.0000 | 0, 0 |
| Stone (1896a) | | | | | |
| S1 | 0.0000 | 0.0000 | 0.0000 | 0.0000 | 0, 0 |
| S2 | 0.0000 | 0.0000 | 0.0000 | 0.0000 | 0, 0 |
| S3 | 0.0000 | 0.0000 | 0.0000 | 0.0000 | 0, 0 |
| S4 | 0.0000 | 0.0000 | 0.0000 | 0.0000 | 0, 0 |
| S5 | -0.0569 | 0.2756 | -0.0047 | 0.1999 | 54, 75 |
| S6 | 0.0000 | 0.0000 | 0.0000 | 0.0000 | 0, 0 |
| S7 | 0.0000 | 0.0000 | 0.0000 | 0.0000 | 0, 0 |
| S8 | 0.0000 | 0.0000 | 0.0000 | 0.0000 | 0, 0 |

| Observation subset: | $\nu_s$ | $\sigma_s$ | $\nu_p$ | $\sigma_p$ | $N_s$, $N_p$ |
|---|---|---|---|---|---|
| Stone (1898c) | | | | | |
| S1 | 0.0000 | 0.0000 | 0.0000 | 0.0000 | 0, 0 |
| S2 | 0.0000 | 0.0000 | 0.0000 | 0.0000 | 0, 0 |
| S3 | -0.1127 | 0.2733 | -0.0086 | 0.2037 | 12, 12 |
| S4 | -0.1519 | 0.2589 | -0.0089 | 0.2315 | 30, 28 |
| S5 | -0.1048 | 0.2048 | 0.0396 | 0.1578 | 15, 15 |
| S6 | 0.0450 | 0.4001 | 0.1994 | 0.2881 | 4, 4 |
| S7 | 0.0000 | 0.0000 | 0.0000 | 0.0000 | 0, 0 |
| S8 | 0.0000 | 0.0000 | 0.0000 | 0.0000 | 0, 0 |
| Morgan (1900) | | | | | |
| S1 | 0.0000 | 0.0000 | 0.0000 | 0.0000 | 0, 0 |
| S2 | 0.0000 | 0.0000 | 0.0000 | 0.0000 | 0, 0 |
| S3 | 0.2297 | 0.2526 | 0.0193 | 0.1762 | 7, 6 |
| S4 | 0.3666 | 0.3065 | 0.0112 | 0.1027 | 6, 4 |
| S5 | 0.5270 | 0.2765 | 0.0594 | 0.1558 | 6, 6 |
| S6 | 0.0000 | 0.0000 | 0.0000 | 0.0000 | 0, 0 |
| S7 | 0.0000 | 0.0000 | 0.0000 | 0.0000 | 0, 0 |
| S8 | 0.0000 | 0.0000 | 0.0000 | 0.0000 | 0, 0 |
| Aitken (1905) | | | | | |
| S1 | 0.0000 | 0.0000 | 0.0000 | 0.0000 | 0, 0 |
| S2 | -0.0412 | 0.2110 | 0.0744 | 0.1370 | 13, 13 |
| S3 | 0.2097 | 0.2004 | 0.0387 | 0.2382 | 13, 12 |
| S4 | 0.2498 | 0.1967 | 0.0827 | 0.2026 | 13, 13 |
| S5 | 0.0000 | 0.0000 | 0.0000 | 0.0000 | 0, 0 |
| S6 | 0.0000 | 0.0000 | 0.0000 | 0.0000 | 0, 0 |
| S7 | 0.0000 | 0.0000 | 0.0000 | 0.0000 | 0, 0 |
| S8 | 0.0000 | 0.0000 | 0.0000 | 0.0000 | 0, 0 |
| Barnard (1910) | | | | | |
| S1 | -0.0255 | 0.0989 | 0.0462 | 0.1269 | 6, 3 |
| S2 | -0.0721 | 0.1794 | 0.0584 | 0.0769 | 18, 8 |
| S3 | 0.0411 | 0.2075 | -0.0668 | 0.0723 | 12, 6 |
| S4 | 0.0642 | 0.1755 | 0.0328 | 0.0913 | 7, 4 |
| S5 | 0.0491 | 0.1116 | 0.1287 | 0.2496 | 8, 6 |
| S6 | 0.0000 | 0.0000 | 0.0000 | 0.0000 | 0, 0 |
| S7 | 0.0000 | 0.0000 | 0.0000 | 0.0000 | 0, 0 |
| S8 | 0.0000 | 0.0000 | 0.0000 | 0.0000 | 0, 0 |
| Aitken (1909) | | | | | |
| S1 | 0.0000 | 0.0000 | 0.0000 | 0.0000 | 0, 0 |
| S2 | -0.0910 | 0.1515 | 0.0621 | 0.0837 | 7, 8 |
| S3 | 0.2377 | 0.2044 | 0.0678 | 0.1255 | 6, 9 |
| S4 | 0.0847 | 0.1824 | 0.0562 | 0.1178 | 8, 8 |
| S5 | 0.0000 | 0.0000 | 0.0000 | 0.0000 | 0, 0 |
| S6 | 0.0000 | 0.0000 | 0.0000 | 0.0000 | 0, 0 |
| S7 | 0.0000 | 0.0000 | 0.0000 | 0.0000 | 0, 0 |
| S8 | 0.0000 | 0.0000 | 0.0000 | 0.0000 | 0, 0 |
| USNO (1929) | | | | | |
| S1 | 0.0000 | 0.0000 | 0.0000 | 0.0000 | 0, 0 |
| S2 | 0.0442 | 0.1108 | 0.0701 | 0.0363 | 2, 2 |
| S3 | 0.0755 | 0.1692 | -0.0128 | 0.1869 | 187, 191 |
| S4 | 0.0521 | 0.1356 | 0.0074 | 0.1511 | 158, 157 |
| S5 | 0.0649 | 0.1719 | 0.0174 | 0.1573 | 318, 320 |
| S6 | 0.0996 | 0.1719 | 0.0106 | 0.1482 | 370, 372 |
| S7 | 0.0000 | 0.0000 | 0.0000 | 0.0000 | 0, 0 |
| S8 | 0.1215 | 0.1606 | -0.0088 | 0.2598 | 117, 120 |

| Observation subset: | $\nu_x$ | $\sigma_x$ | $\nu_p$ | $\sigma_p$ | $N_x$, $N_p$ |
|---|---|---|---|---|---|
| Barnard (1913) | | | | | |
| S1 | 0.1750 | 0.1093 | 0.0451 | 0.0186 | 3, 3 |
| S2 | 0.0726 | 0.1589 | 0.1242 | 0.2092 | 23, 23 |
| S3 | 0.0804 | 0.2407 | 0.0768 | 0.3137 | 45, 44 |
| S4 | 0.0519 | 0.1929 | 0.0711 | 0.3099 | 33, 33 |
| S5 | 0.0718 | 0.2587 | 0.1386 | 0.2770 | 11, 11 |
| S6 | 0.0000 | 0.0000 | 0.0000 | 0.0000 | 0, 0 |
| S7 | 0.0000 | 0.0000 | 0.0000 | 0.0000 | 0, 0 |
| S8 | 0.0000 | 0.0000 | 0.0000 | 0.0000 | 0, 0 |
| Barnard (1915) | | | | | |
| S1 | 0.1383 | 0.0890 | 0.2791 | 0.3066 | 3, 4 |
| S2 | 0.0943 | 0.1738 | 0.0023 | 0.2385 | 13, 13 |
| S3 | 0.1256 | 0.1983 | 0.0580 | 0.2757 | 26, 28 |
| S4 | 0.1291 | 0.2024 | 0.2029 | 0.2359 | 23, 23 |
| S5 | −0.0554 | 0.1118 | 0.0968 | 0.1420 | 11, 11 |
| S6 | 0.0000 | 0.0000 | 0.0000 | 0.0000 | 0, 0 |
| S7 | 0.0000 | 0.0000 | 0.0000 | 0.0000 | 0, 0 |
| S8 | 0.3823 | 0.0000 | 0.1694 | 0.0000 | 1, 1 |
| Barnard (1916) | | | | | |
| S1 | 0.1941 | 0.1928 | 0.0563 | 0.1434 | 13, 13 |
| S2 | 0.0822 | 0.1972 | 0.1071 | 0.2209 | 19, 19 |
| S3 | 0.1344 | 0.1686 | 0.0225 | 0.1796 | 42, 41 |
| S4 | 0.1946 | 0.1835 | 0.0077 | 0.3271 | 21, 20 |
| S5 | 0.1185 | 0.2452 | 0.0232 | 0.2104 | 12, 12 |
| S6 | 0.0000 | 0.0000 | 0.0000 | 0.0000 | 0, 0 |
| S7 | 0.0000 | 0.0000 | 0.0000 | 0.0000 | 0, 0 |
| S8 | 0.0000 | 0.0000 | 0.0000 | 0.0000 | 0, 0 |
| Barnard (1918) | | | | | |
| S1 | −0.0194 | 0.1599 | 0.0703 | 0.0947 | 7, 6 |
| S2 | 0.0826 | 0.2060 | 0.1282 | 0.2910 | 23, 23 |
| S3 | 0.1649 | 0.1274 | −0.0272 | 0.1748 | 36, 35 |
| S4 | 0.1739 | 0.1307 | −0.0074 | 0.2939 | 28, 30 |
| S5 | 0.2478 | 0.1294 | 0.0212 | 0.3503 | 10, 9 |
| S6 | 0.0000 | 0.0000 | 0.0000 | 0.0000 | 0, 0 |
| S7 | 0.0000 | 0.0000 | 0.0000 | 0.0000 | 0, 0 |
| S8 | 0.0000 | 0.0000 | 0.0000 | 0.0000 | 0, 0 |
| Barnard (1927) | | | | | |
| S1 | 0.0125 | 0.2915 | 0.1242 | 0.2063 | 17, 19 |
| S2 | 0.0116 | 0.2274 | 0.0690 | 0.2099 | 65, 66 |
| S3 | 0.1270 | 0.1866 | 0.0215 | 0.1839 | 133, 125 |
| S4 | 0.0936 | 0.1778 | 0.0155 | 0.2015 | 61, 63 |
| S5 | 0.1839 | 0.1887 | 0.0147 | 0.3278 | 45, 43 |
| S6 | 0.0000 | 0.0000 | 0.0000 | 0.0000 | 0, 0 |
| S7 | 0.0000 | 0.0000 | 0.0000 | 0.0000 | 0, 0 |
| S8 | 0.0901 | 0.2952 | 0.0284 | 0.2628 | 8, 8 |
| Struve(1933) Johannesb | | | | | |
| S1 | −0.0016 | 0.1646 | −0.0010 | 0.1896 | 115, 115 |
| S2 | 0.0429 | 0.1265 | −0.0113 | 0.1324 | 187, 190 |
| S3 | 0.0459 | 0.1249 | 0.0309 | 0.1298 | 127, 130 |
| S4 | 0.0331 | 0.1229 | 0.1023 | 0.1136 | 54, 54 |
| S5 | −0.0166 | 0.0268 | −0.0740 | 0.1009 | 2, 2 |
| S6 | 0.0000 | 0.0000 | 0.0000 | 0.0000 | 0, 0 |
| S7 | 0.0000 | 0.0000 | 0.0000 | 0.0000 | 0, 0 |
| S8 | 0.0000 | 0.0000 | 0.0000 | 0.0000 | 0, 0 |

| Observation subset: | $\nu_s$ | $\sigma_s$ | $\nu_p$ | $\sigma_p$ | $N_s$, $N_p$ |
|---|---|---|---|---|---|
| USNO (1954) 61/62 | | | | | |
| S1 | 0.0000 | 0.0000 | 0.0000 | 0.0000 | 0, 0 |
| S2 | 0.0827 | 0.1310 | -0.0510 | 0.1843 | 18, 18 |
| S3 | 0.0478 | 0.2775 | -0.0200 | 0.2730 | 278, 276 |
| S4 | 0.0732 | 0.2193 | -0.0071 | 0.1963 | 283, 284 |
| S5 | 0.0591 | 0.2203 | 0.0175 | 0.2220 | 455, 457 |
| S6 | 0.0572 | 0.1785 | -0.0466 | 0.1737 | 194, 188 |
| S7 | 0.0000 | 0.0000 | 0.0000 | 0.0000 | 0, 0 |
| S8 | -0.0839 | 0.0000 | -0.0360 | 0.0000 | 1, 1 |
| Struve(1933) Yerkes | | | | | |
| S1 | 0.0000 | 0.0000 | 0.0000 | 0.0000 | 0, 0 |
| S2 | -0.1751 | 0.2256 | 0.0512 | 0.2054 | 48, 48 |
| S3 | 0.0000 | 0.0000 | 0.0000 | 0.0000 | 0, 0 |
| S4 | 0.0000 | 0.0000 | 0.0000 | 0.0000 | 0, 0 |
| S5 | 0.0000 | 0.0000 | 0.0000 | 0.0000 | 0, 0 |
| S6 | 0.0000 | 0.0000 | 0.0000 | 0.0000 | 0, 0 |
| S7 | 0.0000 | 0.0000 | 0.0000 | 0.0000 | 0, 0 |
| S8 | 0.0000 | 0.0000 | 0.0000 | 0.0000 | 0, 0 |
| USNO (1954) | | | | | |
| S1 | 0.0000 | 0.0000 | 0.0000 | 0.0000 | 0, 0 |
| S2 | 0.0000 | 0.0000 | 0.0000 | 0.0000 | 0, 0 |
| S3 | 0.0000 | 0.0000 | 0.0000 | 0.0000 | 0, 0 |
| S4 | 0.0000 | 0.0000 | 0.0000 | 0.0000 | 0, 0 |
| S5 | -0.0007 | 0.1547 | 0.0172 | 0.0857 | 32, 32 |
| S6 | 0.0007 | 0.1547 | -0.0172 | 0.0857 | 32, 32 |
| S7 | 0.0000 | 0.0000 | 0.0000 | 0.0000 | 0, 0 |
| S8 | 0.0000 | 0.0000 | 0.0000 | 0.0000 | 0, 0 |
| Harper et al. (1997) | | | | | |
| S1 | 0.0000 | 0.0000 | 0.0000 | 0.0000 | 0, 0 |
| S2 | 0.0000 | 0.0000 | 0.0000 | 0.0000 | 0, 0 |
| S3 | -0.0079 | 0.0377 | 0.0090 | 0.0602 | 184, 184 |
| S4 | -0.0024 | 0.0318 | -0.0053 | 0.0580 | 193, 193 |
| S5 | 0.0016 | 0.0281 | -0.0025 | 0.0545 | 202, 202 |
| S6 | 0.0062 | 0.0213 | -0.0157 | 0.0682 | 118, 118 |
| S7 | 0.0000 | 0.0000 | 0.0000 | 0.0000 | 0, 0 |
| S8 | -0.0023 | 0.0451 | 0.0049 | 0.0708 | 51, 51 |
| Qiao et al. 1999 | | | | | |
| S1 | -0.0448 | 0.1181 | 0.0086 | 0.0174 | 15, 15 |
| S2 | -0.0454 | 0.0802 | 0.0060 | 0.0645 | 47, 47 |
| S3 | -0.0065 | 0.0423 | 0.0050 | 0.0564 | 82, 82 |
| S4 | -0.0037 | 0.0296 | -0.0017 | 0.0423 | 147, 147 |
| S5 | 0.0088 | 0.0368 | -0.0050 | 0.0477 | 151, 151 |
| S6 | -0.0017 | 0.0265 | -0.0076 | 0.0578 | 164, 164 |
| S7 | 0.0000 | 0.0000 | 0.0000 | 0.0000 | 0, 0 |
| S8 | 0.0000 | 0.0000 | 0.0000 | 0.0000 | 0, 0 |
| Veiga et al. (2003) | | | | | |
| S1 | -0.0046 | 0.0534 | -0.0059 | 0.0704 | 329, 329 |
| S2 | 0.0069 | 0.0613 | -0.0038 | 0.0720 | 414, 414 |
| S3 | -0.0041 | 0.0430 | 0.0103 | 0.0710 | 489, 489 |
| S4 | 0.0055 | 0.0522 | -0.0064 | 0.0721 | 527, 527 |
| S5 | -0.0106 | 0.0451 | -0.0106 | 0.0947 | 480, 480 |
| S6 | 0.0293 | 0.0583 | 0.0270 | 0.0932 | 219, 219 |
| S7 | 0.0000 | 0.0000 | 0.0000 | 0.0000 | 0, 0 |
| S8 | -0.0339 | 0.0217 | -0.1685 | 0.0502 | 7, 7 |

| Observation subset: | $\nu_s$ | $\sigma_s$ | $\nu_p$ | $\sigma_p$ | $N_s$, $N_p$ |
|---|---|---|---|---|---|
| Vienne et al. (2001) | | | | | |
| S1 | 0.0439 | 0.0347 | 0.0103 | 0.0345 | 216, 216 |
| S2 | -0.0017 | 0.0421 | 0.0051 | 0.0513 | 860, 860 |
| S3 | -0.0050 | 0.0264 | -0.0052 | 0.0396 | 1747, 1747 |
| S4 | 0.0033 | 0.0247 | 0.0060 | 0.0377 | 1029, 1029 |
| S5 | -0.0018 | 0.0230 | -0.0046 | 0.0507 | 1587, 1587 |
| S6 | 0.0001 | 0.0153 | -0.0006 | 0.0405 | 731, 731 |
| S7 | 0.0000 | 0.0000 | 0.0000 | 0.0000 | 0, 0 |
| S8 | 0.0036 | 0.0291 | 0.0061 | 0.0643 | 523, 523 |
| Harper et al. (1999) | | | | | |
| S1 | 0.0000 | 0.0000 | 0.0000 | 0.0000 | 0, 0 |
| S2 | 0.0000 | 0.0000 | 0.0000 | 0.0000 | 0, 0 |
| S3 | -0.0017 | 0.0272 | -0.0106 | 0.0497 | 245, 245 |
| S4 | -0.0222 | 0.0400 | 0.0114 | 0.1133 | 205, 205 |
| S5 | 0.0055 | 0.0423 | 0.0059 | 0.0856 | 238, 238 |
| S6 | 0.0133 | 0.0351 | -0.0135 | 0.0534 | 242, 242 |
| S7 | 0.0000 | 0.0000 | 0.0000 | 0.0000 | 0, 0 |
| S8 | 0.0076 | 0.0243 | 0.0083 | 0.0383 | 188, 188 |
| Peng, Vienne, Shen (2002) | | | | | |
| S1 | -0.0139 | 0.0302 | 0.0390 | 0.0485 | 54, 54 |
| S2 | -0.0181 | 0.0273 | 0.0062 | 0.0438 | 161, 161 |
| S3 | 0.0093 | 0.0226 | 0.0053 | 0.0281 | 145, 145 |
| S4 | -0.0006 | 0.0182 | -0.0022 | 0.0304 | 161, 161 |
| S5 | 0.0068 | 0.0237 | 0.0055 | 0.0295 | 199, 199 |
| S6 | -0.0010 | 0.0126 | -0.0005 | 0.0505 | 145, 145 |
| S7 | 0.0000 | 0.0000 | 0.0000 | 0.0000 | 0, 0 |
| S8 | -0.0027 | 0.0191 | -0.0281 | 0.0527 | 126, 126 |
| Qiao et al. (2004) | | | | | |
| S1 | -0.0400 | 0.1096 | 0.0466 | 0.1996 | 44, 44 |
| S2 | -0.0682 | 0.1322 | -0.0005 | 0.2055 | 141, 141 |
| S3 | -0.0091 | 0.0866 | 0.0189 | 0.1497 | 236, 236 |
| S4 | -0.0175 | 0.0733 | -0.0230 | 0.0987 | 246, 246 |
| S5 | 0.0174 | 0.0571 | -0.0001 | 0.1237 | 227, 227 |
| S6 | 0.0232 | 0.0401 | -0.0514 | 0.1428 | 207, 207 |
| S7 | 0.0000 | 0.0000 | 0.0000 | 0.0000 | 0, 0 |
| S8 | 0.0291 | 0.0274 | 0.1340 | 0.1958 | 66, 66 |
| Peng et al. (2008) | | | | | |
| S1 | 0.0256 | 0.0712 | 0.0082 | 0.0912 | 358, 358 |
| S2 | -0.0009 | 0.0260 | -0.0003 | 0.0383 | 358, 358 |
| S3 | -0.0051 | 0.0234 | 0.0007 | 0.0296 | 358, 358 |
| S4 | -0.0056 | 0.0237 | -0.0003 | 0.0325 | 358, 358 |
| S5 | -0.0082 | 0.0245 | 0.0021 | 0.0418 | 358, 358 |
| S6 | 0.0074 | 0.0209 | -0.0200 | 0.1202 | 358, 358 |
| S7 | 0.0000 | 0.0000 | 0.0000 | 0.0000 | 0, 0 |
| S8 | 0.0000 | 0.0000 | 0.0000 | 0.0000 | 0, 0 |

Table 3- Correlation between $k_2/Q$ and *da/dt* with all our fitted parameters (Enceladus tidal equilibrium solution), where *a* is the semi-major axis, *l* is the mean longitude, $k=e.cos(\Omega+\omega)$, $h=e.sin(\Omega+\omega)$, $q=sin(i/2).cos(\Omega)$ and $p=sin(i/2).sin(\Omega)$ (with *e* denoting the eccentricity, $\Omega$ denoting the longitude of the node, $\omega$ denoting the argument of the periapsis). Numbers 1,..8 refer to Mimas (S1),..Iapetus (S8), respectively.

|       | $a_1$  | $l_1$  | $k_1$  | $h_1$  | $q_1$  | $p_1$  | $a_2$  | $l_2$  | $k_2$  | $h_2$  | $q_2$  | $p_2$  |
|-------|--------|--------|--------|--------|--------|--------|--------|--------|--------|--------|--------|--------|
| $k_2/Q$ | -0.116 | 0.142 | 0.103 | -0.012 | -0.031 | -0.309 | -0.035 | 0.705 | 0.000 | 0.484 | 0.000 | -0.010 |
| $da/dt$ | 0.085 | -0.020 | -0.125 | -0.044 | 0.082 | 0.375 | 0.027 | -0.588 | -0.005 | -0.415 | 0.011 | 0.008 |

|       | $a_3$  | $l_3$  | $k_3$  | $h_3$  | $q_3$  | $p_3$  | $a_4$  | $l_4$  | $k_4$  | $h_4$  | $q_4$  | $p_4$  |
|-------|--------|--------|--------|--------|--------|--------|--------|--------|--------|--------|--------|--------|
| $k_2/Q$ | -0.043 | 0.169 | 0.014 | -0.037 | -0.199 | 0.219 | -0.170 | 0.372 | 0.088 | -0.070 | 0.036 | -0.027 |
| $da/dt$ | 0.087 | 0.241 | -0.054 | 0.046 | 0.247 | -0.290 | 0.157 | -0.303 | -0.076 | 0.066 | -0.045 | 0.042 |

|       | $a_5$  | $l_5$  | $k_5$  | $h_5$  | $q_5$  | $p_5$  | $a_6$  | $l_6$  | $k_6$  | $h_6$  | $q_6$  | $p_6$  |
|-------|--------|--------|--------|--------|--------|--------|--------|--------|--------|--------|--------|--------|
| $k_2/Q$ | 0.032 | 0.136 | 0.028 | -0.030 | -0.016 | 0.022 | 0.052 | 0.014 | -0.004 | 0.018 | 0.014 | -0.019 |
| $da/dt$ | 0.007 | -0.106 | -0.026 | 0.009 | 0.012 | -0.017 | 0.035 | -0.011 | -0.001 | -0.014 | -0.010 | 0.012 |

|       | $a_7$  | $l_7$  | $k_7$  | $h_7$  | $q_7$  | $p_7$  | $a_8$  | $l_8$  | $k_8$  | $h_8$  | $q_8$  | $p_8$  |
|-------|--------|--------|--------|--------|--------|--------|--------|--------|--------|--------|--------|--------|
| $k_2/Q$ | -0.001 | 0.005 | 0.001 | 0.000 | -0.004 | -0.014 | -0.016 | -0.014 | 0.000 | 0.008 | 0.001 | -0.031 |
| $da/dt$ | 0.001 | -0.004 | -0.002 | 0.000 | 0.003 | 0.009 | 0.023 | 0.013 | 0.001 | -0.024 | -0.002 | 0.024 |

|         | $k_2/Q$ | $da/dt$ |
|---------|---------|---------|
| $k_2/Q$ | 1.000   | -0.838  |
| $da/dt$ | -0.838  | 1.000   |

Table 4- Correlation between all four $k_2/Q$ ratios estimated at the tidal frequency of Enceladus, Tethys, Dione and Rhea, and *da/dt* (Enceladus' eccentricity equilibrium solution).

|            | $k_2/Q(S_2)$ | $k_2/Q(S_3)$ | $k_2/Q(S_4)$ | $k_2/Q(S_5)$ | $da/dt$ |
|------------|--------------|--------------|--------------|--------------|---------|
| $k_2/Q(S_2)$ | 1.000        | 0.020        | -0.197       | 0.003        | -0.012  |
| $k_2/Q(S_3)$ | 0.020        | 1.000        | 0.001        | 0.000        | -0.935  |
| $k_2/Q(S_4)$ | -0.197       | 0.001        | 1.000        | 0.024        | 0.004   |
| $k_2/Q(S_5)$ | 0.003        | 0.000        | 0.024        | 1.000        | 0.006   |
| $da/dt$    | -0.012       | -0.935       | 0.004        | 0.006        | 1.000   |

Table 5- Fitted value of $k_2/Q$ and *da/dt* after removing observation subsets

| Subset removed | $k_2/Q$ | *da/dt* (au/day) | observations |
|---|---|---|---|
| Vienne et al. (2001) | $(2.0 \pm 0.4) \times 10^{-4}$ | $-(11.0 \pm 2.4) \times 10^{-15}$ | 6693 |
| Vass (1997) | $(2.6 \pm 0.4) \times 10^{-4}$ | $-(16.9 \pm 2.3) \times 10^{-15}$ | 3977 |
| FASTT | $(2.6 \pm 0.4) \times 10^{-4}$ | $-(16.9 \pm 2.3) \times 10^{-15}$ | 3137 |
| Struve (1898) | $(4.2 \pm 0.6) \times 10^{-4}$ | $-(20.8 \pm 2.9) \times 10^{-15}$ | 796 |
| USNO (1929) | $(2.6 \pm 0.5) \times 10^{-4}$ | $-(12.9 \pm 2.5) \times 10^{-15}$ | 1162 |

Table A1- Testing the tidal model (tides in the satellites/2-body problem)

| Moon | $\Delta a$ (km) | $\Delta e$ |
|---|---|---|
| Mimas | -5.402801100992870E-002 | -8.537669664267916E-006 |
| Enceladus | -1.587689542119058E-003 | -6.236475859717691E-007 |
| Tethys | -1.229629000688473E-004 | -2.139456833055286E-007 |
| Dione | -8.724264672974581E-005 | -6.140003928510260E-008 |
| Rhea | -1.158774729219763E-005 | -9.617269239444934E-009 |
| Titan | -5.787519520775916E-004 | -8.114568451206283E-009 |